\documentclass[aps,preprint,superscriptaddress,12pt,longbibliography]{revtex4-1}

\usepackage{amsmath}
\usepackage{graphicx}
\usepackage{subcaption}
\usepackage[labelfont=bf]{caption}
\usepackage[colorlinks]{hyperref}
\usepackage{float}
\usepackage{standalone}
\usepackage{import}

\begin{document}

\title{Chain Conformations in Adsorbed Layer during Polymer Capillary Imbibition}

\author{Tao Liang}
\affiliation{South China Advanced Institute for Soft Matter Science and Technology, School of Emergent Soft Matter, South China University of Technology, Guangzhou 510640, China}
\affiliation{Guangdong Provincial Key Laboratory of Functional and Intelligent Hybrid Materials and Devices, South China University of Technology, Guangzhou 510640, China}

\author{Li Peng}
\affiliation{National-Certified Enterprise Technology Center, Kingfa Science and Technology Co., LTD., Guangzhou 510663, China}

\author{Xianbo Huang}
\email[]{huangxianbo@kingfa.com.cn}
\affiliation{National-Certified Enterprise Technology Center, Kingfa Science and Technology Co., LTD., Guangzhou 510663, China}

\author{Jiajia Zhou}
\email[]{zhouj2@scut.edu.cn}
\affiliation{South China Advanced Institute for Soft Matter Science and Technology, School of Emergent Soft Matter, South China University of Technology, Guangzhou 510640, China}
\affiliation{Guangdong Provincial Key Laboratory of Functional and Intelligent Hybrid Materials and Devices, South China University of Technology, Guangzhou 510640, China}


\begin{abstract}
We conducted molecular dynamics simulations to investigate chain conformations in adsorbed layers during polymer capillary imbibition.
While the imbibition length adheres to the classical Lucas-Washburn equation, a notable deviation in mobile bead density emerges under strong confinement, consistent with \emph{in situ} dielectric spectroscopy experiments.
The proportion of loop structures within adsorbed layers progressively increases during capillary infiltration, attributed to the relaxation of initially stretched chains toward equilibrium configurations.
Furthermore, systematic analysis revealed that chain relaxation dynamics exhibit length-dependent retardation, especially under high confinement.
The characteristic desorption time demonstrates chain-length dependence in quantitative agreement with scaling predictions.

\end{abstract}


\maketitle

\newpage
\section{Introduction}
Understanding capillary infiltration of polymers is important in many aspects, 
including nanotechnology \cite{Stukalin_2013, van_der_Kooij_2017, Wang_2020}, separation of biomacromolecules \cite{Bakry_2009} and lab-on-chip device \cite{Olanrewaju_2018}. The capillary filling of simple liquids can be well described by the classic Lucas-Washburn equation (LWE) \cite{Lucas_1918, Washburn_1921}:
\begin{equation}
	h(t)=\sqrt{\frac{\gamma R\cos\theta}{2\eta}} t^{1/2}
	\label{LWE}
\end{equation}
where $h(t)$ is the time-dependent infiltration length, $R$ is the radius of capillary tube, $\gamma$, $\theta$ and $\eta$ are the surface tension, contact angle and viscosity of the fluid, respectively.

A basic assumption of LWE is that the size of fluid particles is much less than the capillary radius, thus the fluid can be treated as a continuous medium.
This assumption is valid for simple fluids, for example, water (size $10^{-10}$ m) filling into mm-size glass tubes. However, when it comes to the infiltration of polymer melt ($R_g \approx 10$-$100$ nm) into self-ordered nanoporous aluminum oxide (AAO) (also $R=10$-$100$ nm), there are noticeable deviations from the LWE prediction.
The infiltration length $h(t)$ still follows the $t^{1/2}$ dependence, but the prefactor varies depending on the ratio between polymer chain size and capillary tube radius.
Recent experiment \cite{Yao_2017} reported a reversal in the dynamics of capillary filling with polymer molecular weight. Polymer melt of short chains shows a smaller prefactor than the LWE equation, while polymer melt of longer chains displays a larger prefactor.
Later, Yao \textit{et al.} \cite {Yao_2018} proposed a unified theory to explain the reversal by demonstrating two competing mechanisms. One is the adsorption of chains on the surface of capillary tube, which creates a dead-layer and reduces the effective tube radius, resulting a higher effective viscosity. The other is the reptation of chains under confinement, which allows the melt to exhibit a lower effective viscosity due to the disentanglement.
Using molecular dynamics simulations \cite{Zhang_2023}, we have confirmed the underlying molecular pictures proposed in Ref. \cite{Yao_2018}.
Yao \textit{et al.} \cite{Yao_2018a} also found an enrichment of the short chains inside the nanopores and near the pore surface by exploring the capillary imbibition of binary blends consisting of short and long chains. 

The reversal occurs not only in capillary tubes with well-defined geometry, but also in irregular geometries, such as porous media created by nanoparticle packing. The infiltration of unentangled polymer melt into nanoparticle packing slows down as the ratio between polymer chain size and average pore size (confinement ratio) increases \cite{Hor_2018, Hor_2018_softmatter, Venkatesh_2021}. Meanwhile, the infiltration of entangled polymer melt into nanoparticle packing enhances as the confinement ratio increases \cite{Venkatesh_2022, Venkatesh_2022_a}.
In a more recent work, Heo \textit{et al.} \cite{Heo_2024} demonstrated that adsorbed layers could play an important role in polymer capillary filling. They explored the infiltration dynamics of poly(styrene-\textit{stat}-2-vinyl-pyridine) (PS-\textit{stat}-P2VP) statistical copolymers undergoing capillary filling into the interstices of silica nanoparticle packings ($\text{SiO}_2$ NP). They observed a non-monotonic dependence of the time required to fully fill $\text{SiO}_2$ NP on the fraction of 2VP. They ruled out the influence of other properties like zero-shear viscosity, glass transition temperature, \textit{et al.}, and attributed the non-monotonic dependence to the adsorbed layers. The above results have shown that the adsorption of polymer chains on nanopore surfaces warrants deeper investigation.   

Previously, the imbibition length was monitored by \textit{ex situ} reflection microscopy \cite{Yao_2017}, which is time-consuming and lacks molecular information.
The so-called \textit{nanofluid} has been widely applied in investions of polymer properties under confinement \cite{Serghei_2010epl, Serghei_2010small, Serghei_2010softmatter, Serghei_2013}.
In this method, two surfaces of AAO templates perpendicular to the infiltration direction, are sputtered with a thin gold layer.
This operation transforms the AAO templates to capacitors when an AC field is applied along the pore axes, and the dielectric signals are collected.
Because dielectric strength of segmental process is proportional to the number density of dipoles. Tu \textit{et al.} found the  dielectric strength increases linearly with the square root of time by \textit{in situ} monitoring time-dependent imbibition length $h(t)$ of PnBMA in nanopores \cite{Tu_2019}.
For a different polymer cis-1,4-polyisoprene (cis-PI), which has stronger interaction with AAO surface than PnBMA did, the result is quite different \cite{Tu_2020}.
In large and small nanopores, the dielectric strength of segmental process increases linearly with the squart root of time initially. But it slows down for large nanopores and even kept constant for small nanopores at later times.
Tu \textit{et al.} constructed a plausible microscopic scenario to explain the experiment results \cite{Tu_2020}.
In early stage of capillary filling, relative fewer chains are adsorbed on the surface of capillary tube, and the adsorbed chains contain a small fraction of loop structures. However, in later stage, more chains are adsorbed and the adsorbed chains contain a large fraction of loop structures. Chains that are not initially adsorbed on the surface become adsorbed onto the surface during capillary filling process. 

With \textit{in situ} dielectric spectroscopy, Tu \textit{et al.} \cite{Tu_2021} also measured the adsorption kinetics after full imbibition by following the evolution of the dielectrically active longest normal mode. The dielectric strength of longest normal mode continues to decrease after full imbibition, which provides a way to characterize the adsorption kinetics of chains. The results show that adsorption characteristic time strongly depends on the ratio $R_g/R$.
For a given pore radius, the molar dependence of characteristic adsorption time is in good agreement with a scaling theory developed by de Gennes \cite{de_Gennes_1987}. 

Previous studies have demonstrated the importance of adsorbed layers, but due to experimental difficulties, a comprehensive picture at molecular scale is still missing.
In this work, we use molecular dynamics simulations to study the conformations of chains adsorbed on the tube surface.
As the experimental results have shown, the chains on the capillary surface change conformation during and after capillary filling, our simulation results provide a detailed anlysis on this change.
We have studied systems with different chain lengths and capillary tube radii, confirming the increment in the loop fraction of adsorbed chains and observing a stretched-relaxed transition of adsorbed chains.  

The article is organized as follows:
In the "Simulation Model" section, we introduce the simulation model and some basic definitions, such as adsorbed repeat units, mobile repeat units, the time of full infiltration, etc.
In the "Results and Discussions" section, we present the simulation results, including the conformations of adsorbed chains during capillary filling, the relaxation dynamics of chains, and the adsorption kinetics after full infiltration.
Lastly, we provide the conclusion of the article in the "Summary" section.

\section{Simulation Model}
\begin{figure}[ht]
	\centering
	\includegraphics[scale=0.5]{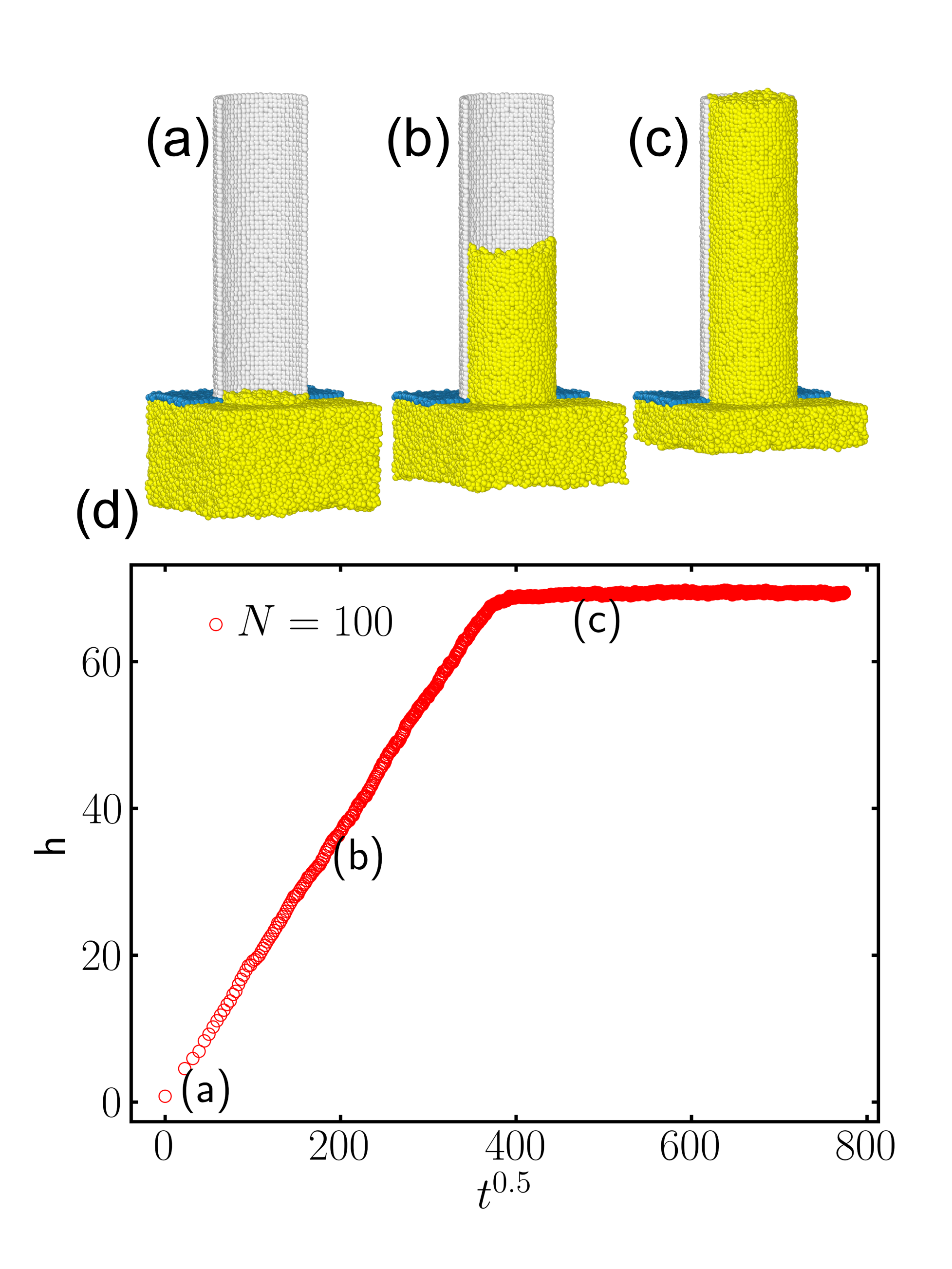}
	\caption{Snapshots of our simulation model of the overall infiltration processes in system with radius $R=10.85\,\sigma$ and chain length $N=100$. (a), (b) and (c) are snapshots corresponding to three time points that are labeled in (d). (d) The capillary infiltration height $h(t)$ as a function of $t^{0.5}$.}
	\label{model}
\end{figure}

We create a simulation box of size $40\sigma\times40\sigma\times200\sigma$ in $x, y, z$ directions, as shown in Figure \ref{model}(a), where $z$-direction is parallel to the tube's axis. A single layer of face-centered cubic lattice beads (with a lattice constant $2.467\sigma$ to prevent polymer melt from passing through), referred to as W, servers as a reservoir perpendicular to $z$-axis at $(0,0,0)$. Just above the center of the reservoir, we construct a cylindrical tube named T, along the $z$-axis. Cylindrical tubes with different capillary radii ($R=5.85,10.85,15.85\,\sigma$) are formed by a single layer of square lattice beads, with a lattice spacing of $1.08\,\sigma$. The tube height is set to $H=70\,\sigma$. W beads within capillary radius $R$ is removed. All W and T beads are bonded to virtual beads at their initial position using a bond potential $U(r)=\frac{1}{2}kr^2$, where $r$ is the bond length. This allows W and T beads to fluctuate around their initial positions, corresponding to the thermostat temperature \cite{Dimitrov_2007, Zhang_2023, Zhang_star}. The nonbonded potential between all beads are described by the truncated-shifted Lennard-Jones (LJ) potential as equation \ref*{lj_interaction} with a cutoff distance $r_{\mathrm{cut}}=2.5\sigma$ and $\sigma=0.8$ \cite{Dimitrov_2007}. The strength of nonbonded interactions between W and T is set to be $\epsilon_{\mathrm{TT}}=\epsilon_{\mathrm{TW}}=\epsilon_{\mathrm{WW}}=1.0\,k_BT$, ensuring that polymer melt beads cannot pass through the W and T bead walls.
\begin{equation}
	U\left( r \right) =\begin{cases}
		4\epsilon_{\mathrm{ij}}\left[ \left( \frac{\sigma}{r_{\mathrm{ij}}} \right) ^{12}-\left( \frac{\sigma}{r_{\mathrm{ij}}} \right) ^6-\left( \frac{\sigma}{r_{\mathrm{cut}}} \right) ^{12}+\left( \frac{\sigma}{r_{\mathrm{cut}}} \right) ^{6} \right] &,r\leq r_{\mathrm{cut}}\\
		0&,r>r_{\mathrm{cut}}\\
	\end{cases}
	\label{lj_interaction}
\end{equation} 
We study five types of polymer chains named A with lengths $N=50, 100, 200, 300, 400$. The bonded potential of polymer chains are described by the finite extension nonlinear elastic (FENE) potential \cite{Kremer_1990}. 
\begin{equation}
	U_{\mathrm{FENE}}=-\frac{1}{2}k_{\mathrm{spring}}R_{\mathrm{max}}^2\ln{(1-\frac{r^2}{R_{\mathrm{max}}^2})}
\end{equation}
where $k_{\mathrm{spring}}=30\,k_BT/\sigma^2$ and $R_{\mathrm{max}}=1.5\,\sigma$.  

For capillary radii $R=5.85\,\sigma$ and $R=10.85\,\sigma$, the total number of polymer beads is 36000,  while for $R=15.85\,\sigma$, it is 72000. The strength of nonbonded interactions  between polymer beads is set to $\epsilon_{\mathrm{AA}}=1.4\,k_BT$. Before capillary infiltration, we set the interaction strengths as $\epsilon_{\mathrm{AW}}=1.4\,k_BT$ and $\epsilon_{\mathrm{AT}}=0.5\,k_BT$ to allow the polymer melt to relax under the reservoir wall.
After sufficient relaxation, we increase $\epsilon_{\mathrm{AT}}$ from $0.5 \,k_BT$ to $1.6\,k_BT$, initiating the infiltration process. We employ the dissipative particle dynamics (DPD) thermostat \cite{Espa_ol_1995, Groot_1997, Junghans_2008}, with a temperature of $T=1.0$ and a friction parameter of $\xi=0.5$. The integration time step is set to $\Delta t=0.005\,\tau$. Our simulations are conducted using HOOMD-blue package \cite{Anderson_2020}.

Since our goal is to study the chain conformations of the adsorbed layer, we first need to define the adsorbed layer. Previous studies \cite{Smith_2004, De_Virgiliis_2012, Zhou_2023} have shown that repeat units near planar surfaces exhibit a layered distribution, which can be characterized by repeat units density profile. 
The first layer of repeat units near an attractive substrate is typically considered the adsorbed layer. 
We calculate the repeat units density profiles in systems with all three tube radii (as shown in Figure S1), which all show layered distribution. Based on this, we define repeat units within the first layer as adsorbed repeat units, and chains containing at least one adsorbed repeat unit as adsorbed chains. Furthermore, we classify repeat units in all chains inside tube into four types:
\begin{itemize}
\item {\bf Train}: adsorbed repeat units in direct contact with the surface.
\item {\bf Loop}: non-adsorbed repeat unit sequences located between train segments.
\item {\bf Tail}: non-adsorbed repeat unit sequences at the ends of adsorbed chains.
\item {\bf Free}: chains with no repeat units in the adsorbed layers.
\end{itemize}

Because we need to analyze polymer chain conformations during and after capillary filling, we need to determine the time $t^*$ of full infiltration.
The imbibition length $h(t)$, as a linear function of square root of time $t^{1/2}$, is fitted by two trend lines: one during imbibition and the other after \cite{Heo_2024}.
Then, the time $t^*$ required to fully fill the capillary tube is determined by finding the intersection of two trend lines (the result is shown in Figure S2). 

\section{Results and Discussions}
\subsection{During Capillary Filling}
As train structures are adsorbed on the capillary tube surface and the two ends of loop structures are fixed by train repeat units, they do not fluctuate over time and therefore make no contribution to the dielectric strength. To compare with experimental results of dielectric strength, we define mobile repeat units as those that include tail structures in the adsorbed layers and free chains. We calculate the amount of mobile repeat units $N_{\rm mobile}$ during and after capillary infiltration. 
In capillary filling of PnBMA, early experimental results \cite{Tu_2019} have shown that dielectric strength $\Delta\epsilon(t)$ of the polymer melt inside the capillary tube increases linearly with $t^{1/2}$.  
However, in capillary filling of cis-PI 42k [23], the increase in dielectric strength slows down with capillary tube radius $R=20$ \text{nm} (confinement ratio $R_g/R=0.35$). But the slowing down phenomenon is less obvious with capillary tube radius $R=50$ \text{nm} (confinement ratio $R_g/R=0.14$).

The amount of mobile repeat units $N_{\mathrm{mobile}}$ as a function of $t^{1/2}$ is shown in Figure 2a. In our simulation systems with tube radius $R=5.85\,\sigma$, as chain lengths range from $N=50$ to $N=400$, $R_g/R$ ranges from $0.56$ to $1.43$. The confinement ratio ($R_g/R$) used in the simulations is comparable to that in the experiments.
As chain lengths increase, $N_{\mathrm{mobile}}(t^{1/2})$ gradually shows two time regimes, with the second one being slower than the first one. 
The slowing down effect in the growth of mobile repeat units is more evident as chain lengths increase. This can be explained by the increase of loop ratio and decrease of tail ratio as chain length increase (as shown in Figure S3).  
However, it is difficult to identify the same tendency when capillary tube radii increase (the results are shown in Figure S4 and Figure S5). These results are consistent with experimental results \cite{Tu_2020} when capillary tube is wider. When the radius of capillary tube increases, the ratio between surface and volume becomes smaller. The fraction of repeat units in adsorbed layers gradually decreases, so the adsorbed layers have less influence on the amount of mobile repeat units. Thus, the slowing down phenomenon is less obvious.
\begin{figure}[htp]
	\centering
	\begin{subfigure}[ht]{0.5\textwidth}
		\centering
		\includegraphics[width=\textwidth]{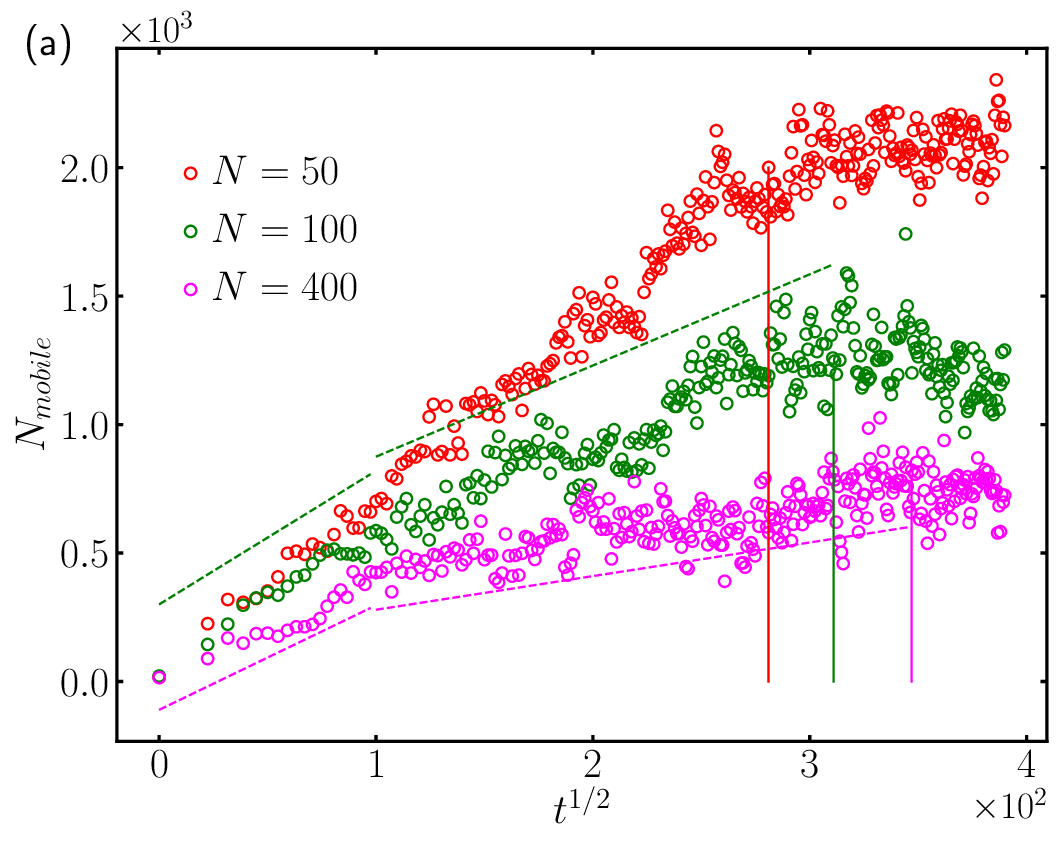}
	\end{subfigure}
	\vspace{0cm}
	\begin{subfigure}[ht]{0.5\textwidth}
		\centering
		\includegraphics[width=\textwidth]{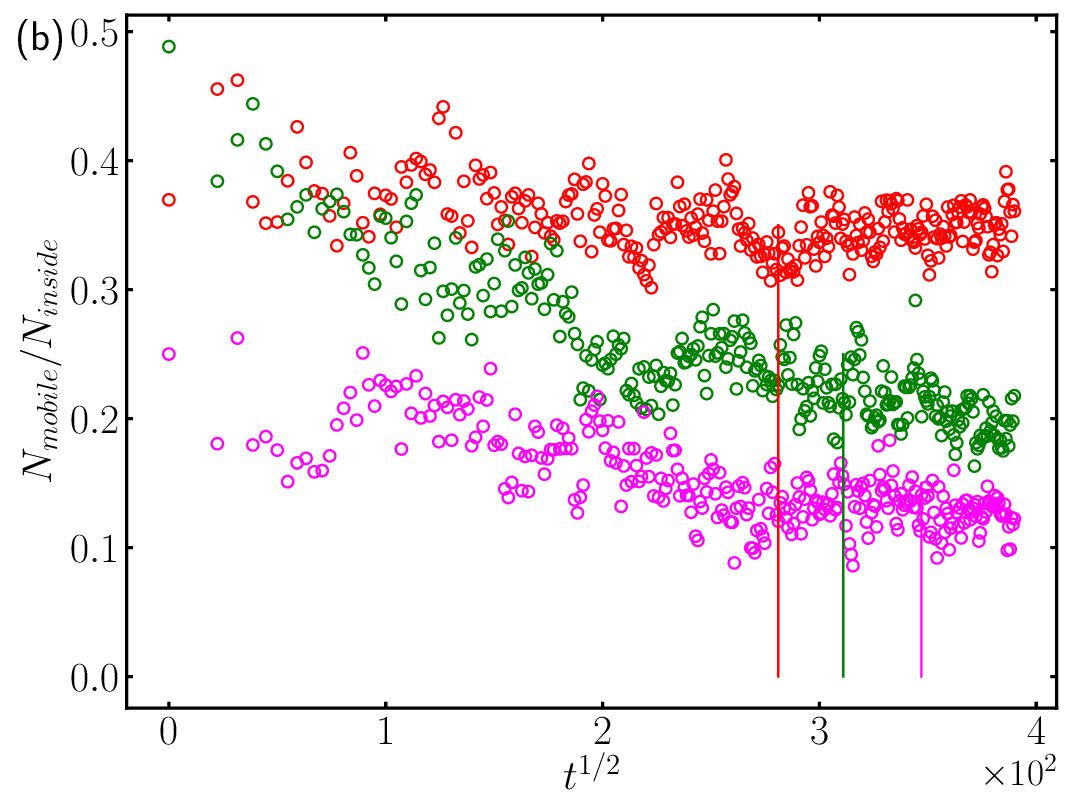}
	\end{subfigure}
	\caption{The amount (a) and the fraction (b) of mobile repeat units versus $t^{1/2}$ with capillary radii $R=5.85\sigma$, polymer melt of length $N=50$ is shown in red, $N=100$ is shown in green and $N=400$ is shown in purple. The dashed lines show tendency at $t^{0.5}\in(0, 100)$ and $t^{0.5}\in(100, t^*)$. The vertical solid  lines indicate the time $t^*$ of full infiltration.}
	\label{mobile_5}
\end{figure}
The fraction of mobile repeat units among all repeat units inside the tube during infiltration process with capillary radius $R=5.85\sigma$ is shown in Figure \ref{mobile_5}(b).
The fractions of three kinds of chain lengths decrease during infiltration, which are consistent with the slowing down of $N_{\mathrm{mobile}}$.  

\begin{figure}[ht]
	\centering
	\begin{subfigure}[ht]{0.5\textwidth}
		\centering
		\includegraphics[width=\textwidth]{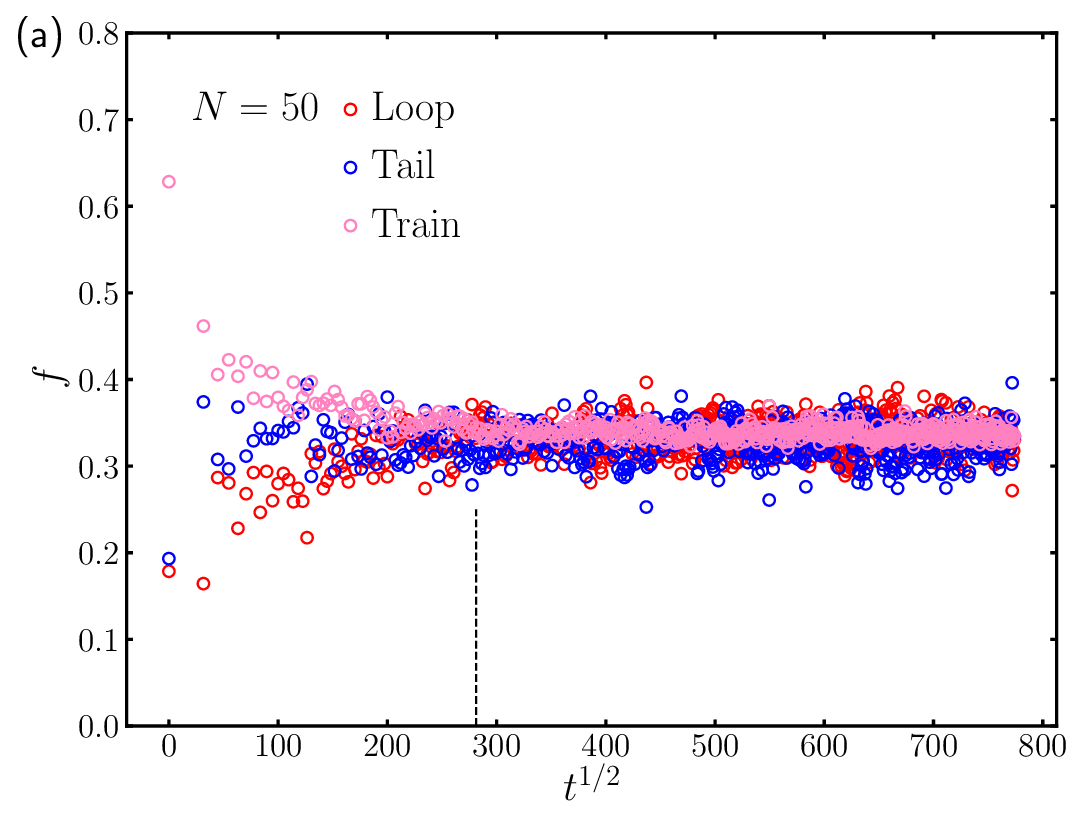}
		\label{ratio_R5_50}
	\end{subfigure}
	\vspace{0cm}
	\begin{subfigure}[ht]{0.5\textwidth}
		\centering
		\includegraphics[width=\textwidth]{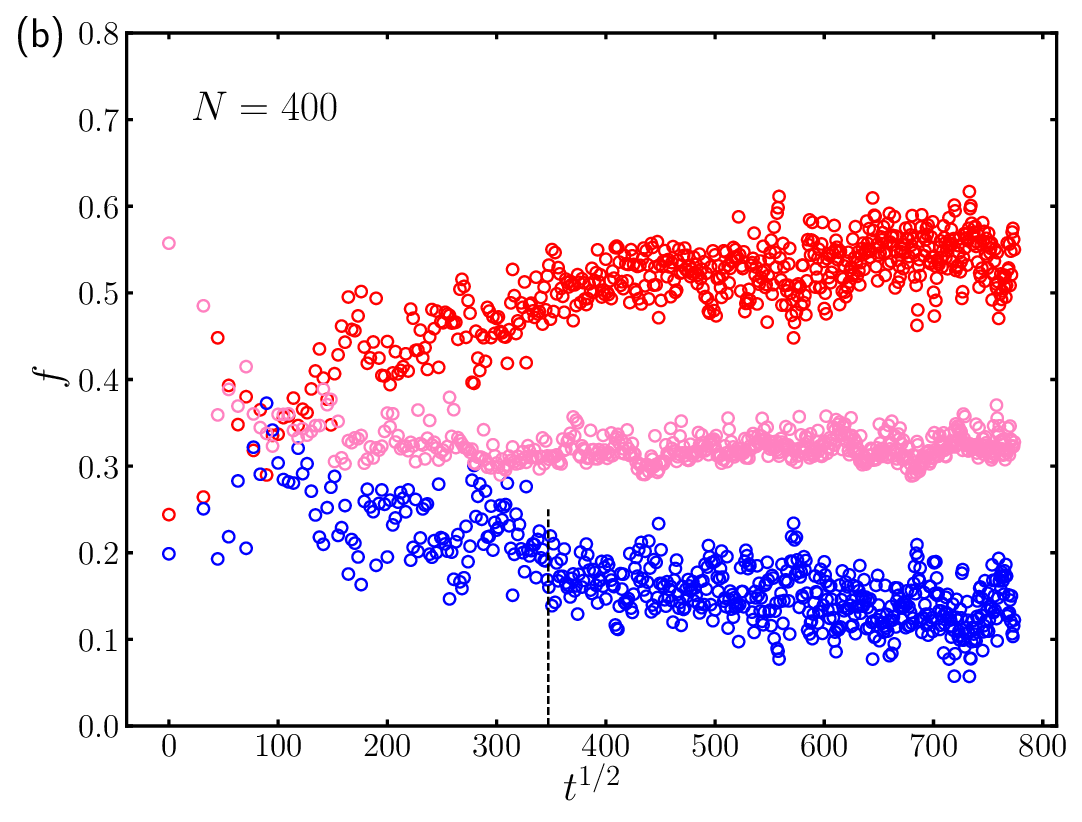}
		\label{ratio_R5_400}
	\end{subfigure}
	\vspace{0cm}
	\caption{Ratio of three types structure of adsorbed monomers (loop shown in red, train shown in pink and tail shown in blue) with capillary tube radii $R=5.85\sigma$. (a): chain length $N=50$, (b): chain length $N=400$. The vertical dash lines indicate the time when capillary tube is full filled.}
	\label{ratio_R5}
\end{figure}

For polymer conformations in the adsorbed layers, we calculate the ratios of three types of repeat units (train, loop and tail) during capillary infiltration. The result with a radius $R=5.85\,\sigma$ is shown in Figure \ref{ratio_R5}. In order to calculate the change in polymer conformations in early stages, we calculate the ratio of three types of repeat units when part of chain is inside the tube and adsorbed on surface. Thus, we calculate the ratio of different types of repeat units ($f_{\mathrm{train}}, f_{\mathrm{loop}}, f_{\mathrm{tail}}$) for the inner part of adsorbed chains. Figure \ref{ratio_R5} shows that, independent of chain length, $f_{\mathrm{loop}}$ increases in early stage of capillary infiltration and $f_{\mathrm{train}}$ decreases simultaneously.

\begin{figure}[ht]
	\centering
	\begin{subfigure}[ht]{0.5\textwidth}
		\centering
		\includegraphics[width=\textwidth]{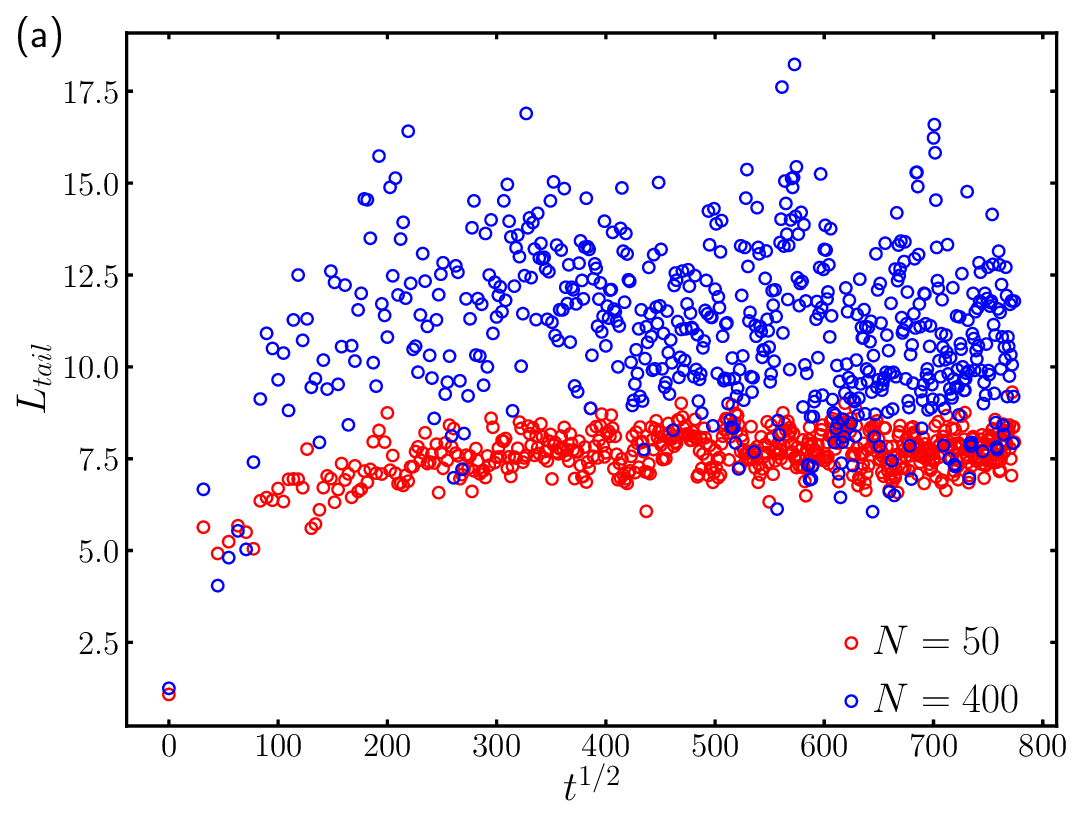}
		\label{In_Tail_R5_tail}
	\end{subfigure}
	\vspace{0cm}
	\begin{subfigure}[ht]{0.5\textwidth}
		\centering
		\includegraphics[width=\textwidth]{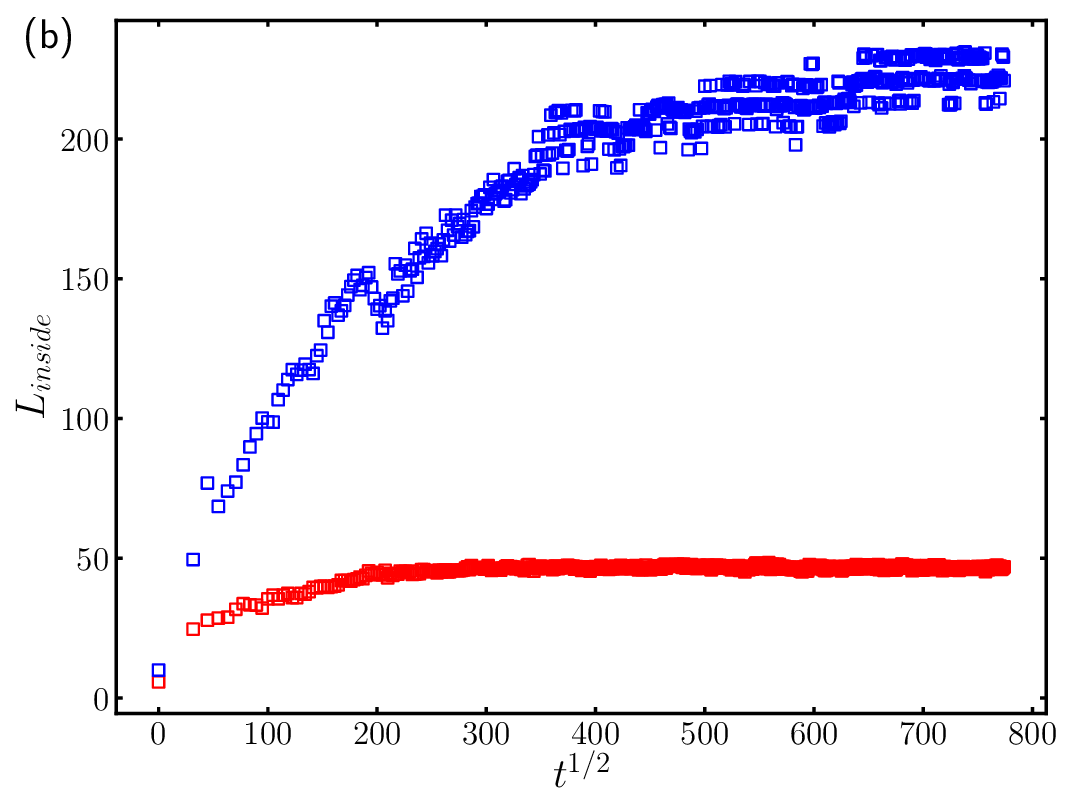}
		\label{In_Tail_R5_in}
	\end{subfigure}
	\caption{(a): Average length of tail structure of adsorbed chains $L_{\mathrm{tail}}$ during capillary infiltration. (b): Average length of part of adsorbed chains that are inside the capillary tube $L_{\mathrm{inside}}$ during capillary infiltration. Red: chain length $N=50$, blue: chain length $N=400$.}
	\label{In_tail_R5}
\end{figure}

For tail type of monomers specially, $f_{\mathrm{tail}}$ for chain length $N=50$ almost remains constant during capillary filling while $f_{\mathrm{tail}}$ for chain length $N=400$ gradually decreases simultaneously.
In Figure \ref{In_tail_R5}, tail lengths ($L_{\mathrm{tail}}$) for both chain lengths $N=50$ and $N=400$ initially increase, then fluctuate around a constant value. 
We also calculate the length of chains that are both inside the tube and some repeat units of the chain are adsorbed on tube surface, donated as $L_{\mathrm{inside}}$. $L_{\mathrm{inside}}$ in system of chain length $N=50$ increases in the same way as tail length. Therefore, the ratio of tail type of repeat units remains constant. However, $L_{\mathrm{inside}}$ for chain length $N=400$ continues increasing during capillary infiltration because it takes more time for the whole chain to get inside the tube, which can explain the decrease of $f_{\mathrm{tail}}$ for chain length $N=400$. 
Thus, we can attribute the increase in $f_{\mathrm{loop}}$ to the decrease in $f_{\mathrm{train}}$.
This result can be observed in systems with all radii and chain lengths (as shown in Figure S7), which turns out to be a universal phenomenon.

\begin{figure}[ht]
	\centering
	\includegraphics[scale=0.5]{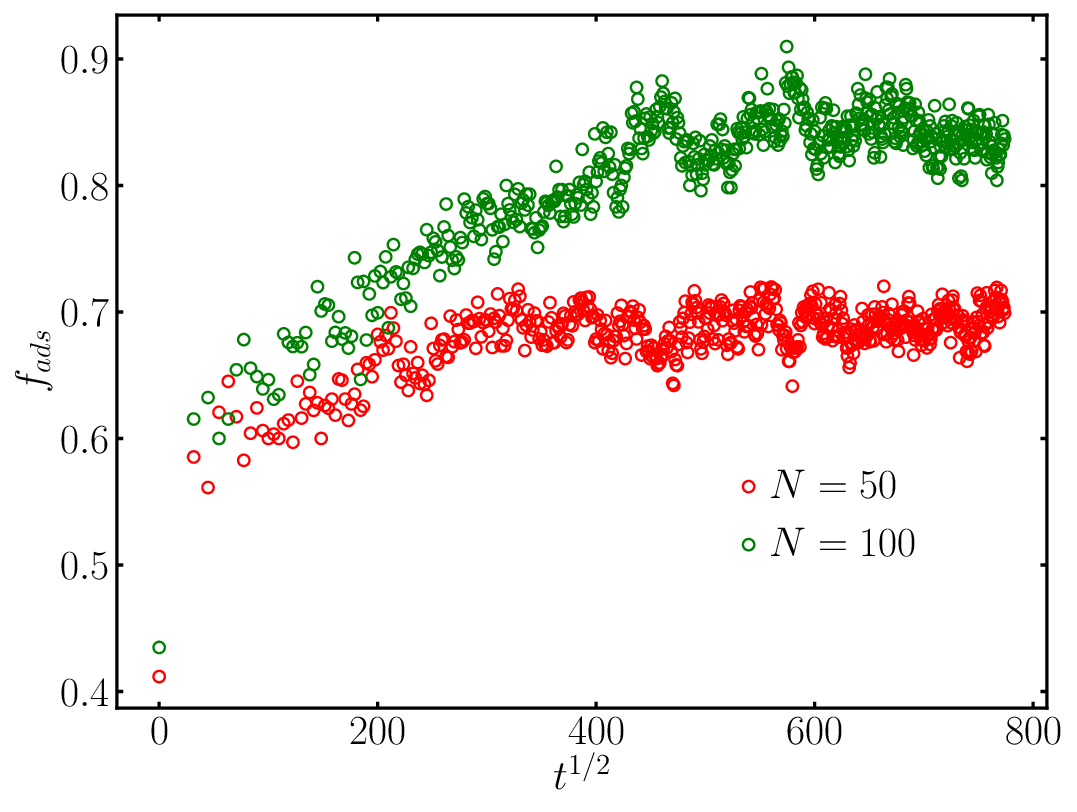}
	\caption{The ratio between the amount of adsorbed chains and the amount of chains that are inside the tube in system with tube radius $R=10.85\,\sigma$ and chain length $N=50,100$.}
	\label{r_ads_R10_1}
\end{figure}

We calculate the ratio of adsorbed chains $f_{\mathrm{ads}}$ among all chains inside the tube in system with radius $R=10.85$. The result in Figure \ref{r_ads_R10_1} shows that $f_{\mathrm{ads}}$ increases with time. 
The experimental result in spectrum of normal modes \cite{Tu_2021} has show the
narrowing of the distribution of relaxation times with time.
The author contribute the broader distribution at the earlier stages to chain adsorptions. 
The result in Figure \ref{r_ads_R10_1} is consistent with experimental result. 

Our simulation model can reproduce the experimental results, as shown by the amount of mobile repeat units. The slowing down of the increase in $N_{\mathrm{mobile}}$ in system with capillary tube radius $R=5.85\,\sigma$ is consistent with the slowing down of dielectric strength in capillary with $R=20\,\text{nm}$. During capillary filling, the increase in $f_{\mathrm{loop}}$ and $f_{\mathrm{ads}}$ can confirm the assumption developed by Tu \textit{et al.} \cite{Tu_2020}.       

\begin{figure}[htp]
	\centering
	\begin{subfigure}[ht]{0.5\textwidth}
		\centering
		\includegraphics[width=\textwidth]{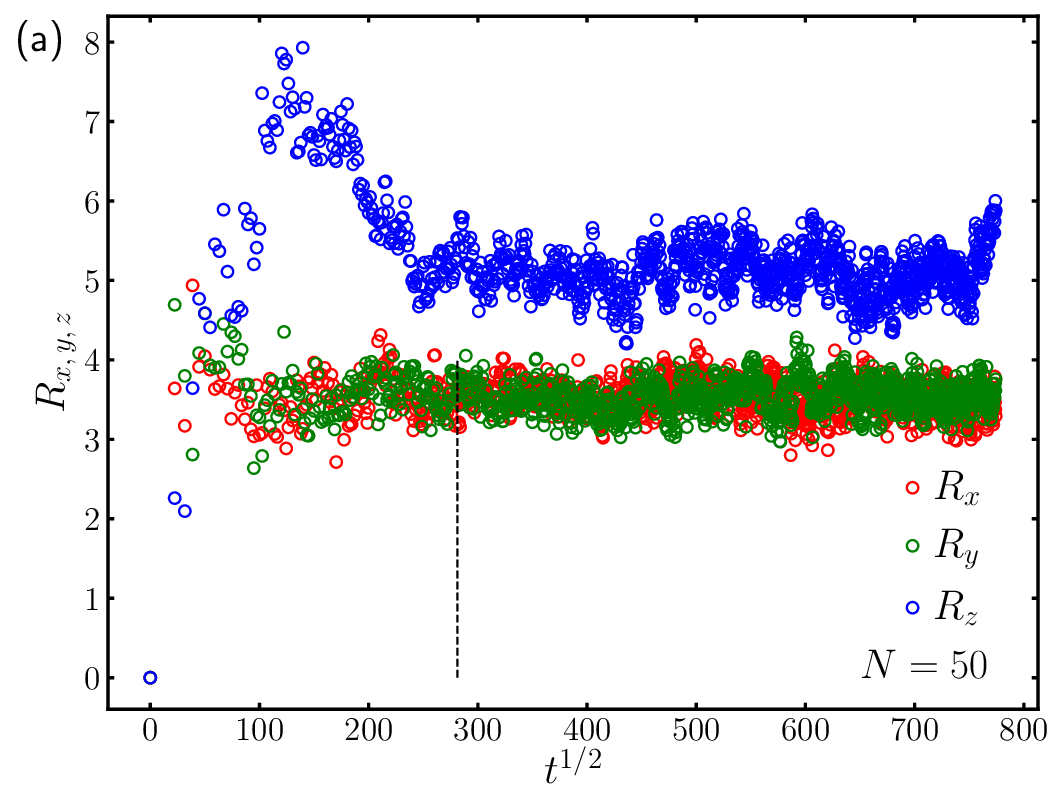}
		\label{R5xyz_N50}
	\end{subfigure}
	\vspace{0cm}
	\begin{subfigure}[ht]{0.5\textwidth}
		\centering
		\includegraphics[width=\textwidth]{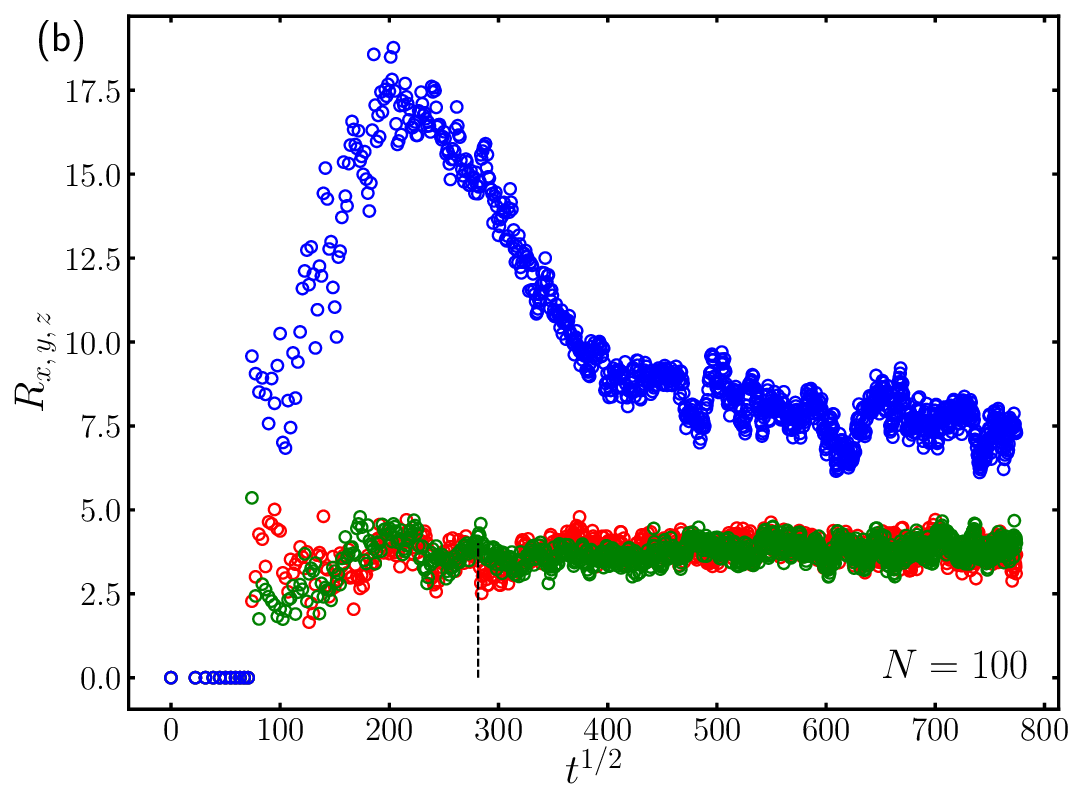}
		\label{R5xyz_N100}
	\end{subfigure}
	\vspace{0cm}
	\caption{End-to-end distances in \textit{x, y, z} directions $R_{x,y,z}$ during capillary infiltration with tube radius $R=5.85\,\sigma$. Chain length $N=50$ (a), $N=100$ (b). The vertical dash lines indicate the time when capillary tube is fully filled.}
	\label{R5xyz}
\end{figure}

In order to look deeper into the conformations of adsorbed layers, we also measure the end-to-end distance $R_{\mathrm{ee}}$ of adsorbed chains in three directions (\textit{x, y, z}) during capillary infiltration. Unlike the calculation of three types repeat units, here we only consider chains that are entirely inside the tube and have at least one repeat unit adsorbed on the surface. Here we present the results for systems with radius $R=5.85\,\sigma$ and chain lengths $N=50, 100$, as shown in Figure \ref{R5xyz}. We can observe $R_z$ first increases and then decreases to a constant value. The initial increase in $R_z$ is due to the low contact angle ($\cos\theta>0.8$ \cite{Zhang_2023}) between polymer melt and capillary tube surface in the condition of $\epsilon_{AT}=1.6\,k_BT/\sigma$. This phenomenon is similar with adsorption of polymer onto planar surface \cite{Linse_2010, K_llrot_2010, K_llrot_2009, De_Virgiliis_2012, Behbahani_2019}. When chains initially adsorbed on planar surface, they were stretched parallel onto the surface and then relax towards equilibrium states after fully adsorbed. These two different processes can explain the initial increase in $R_z$ followed by its subsequent decrease. Both $R_x$ and $R_y$ remain constant during capillary infiltration, as the driving force acts along the \textit{z}-direction. We also calculate the results in systems with radius $R=5.85\,\sigma$ and longer chain lengths $N=200, 300, 400$. It is difficult to observe the same phenomenon as it takes more time for chains to be entirely inside the tube (as shown in Figure S8).  

\begin{figure}[ht]
	\centering
	\includegraphics[scale=0.5]{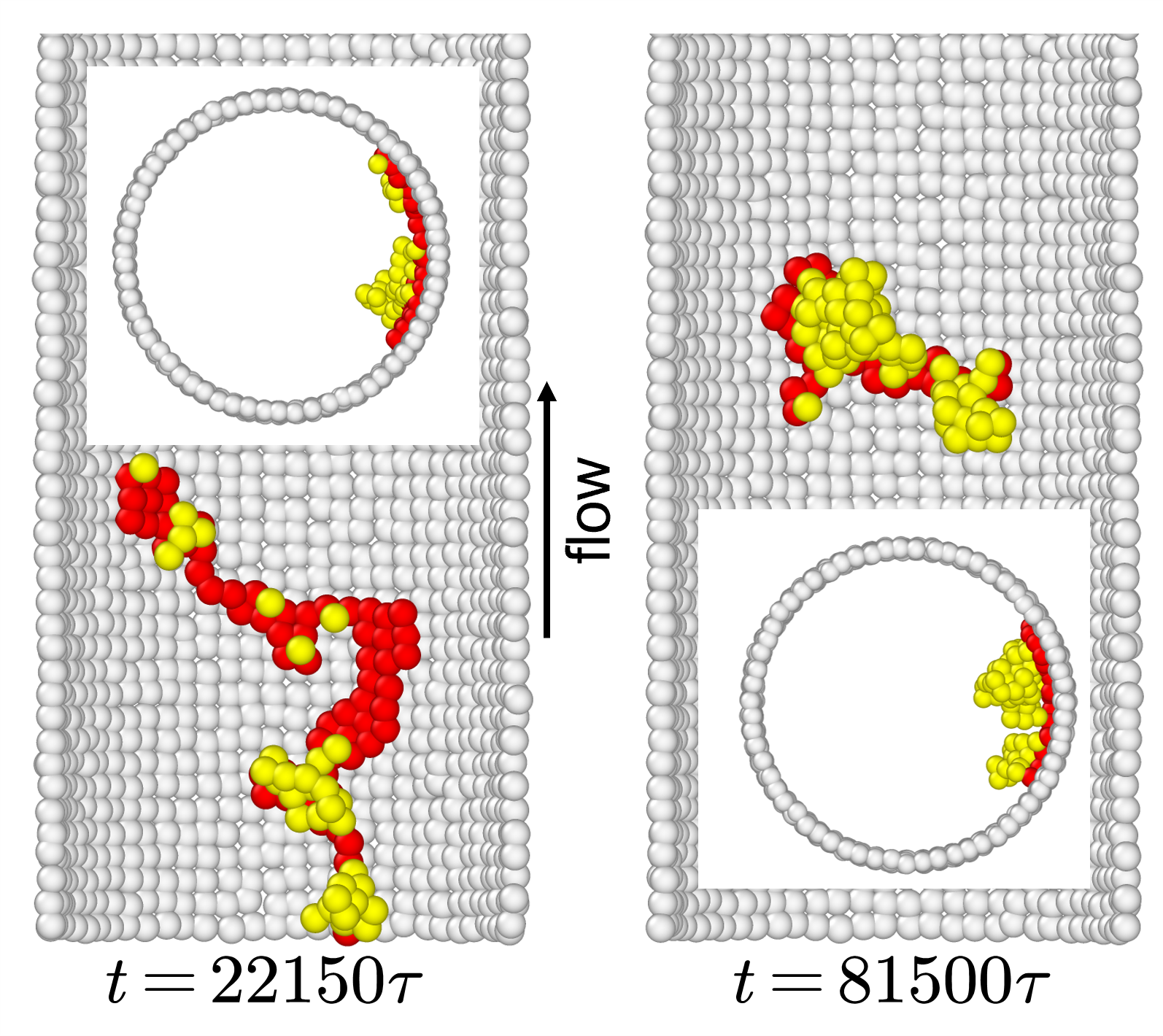}
	\caption{Snapshot of one adsorbed chain at different times. The main and insert images show views along and perpendicular to the cylinder axis, respectively. Tube beads are shown in white, adsorbed beads in red and non-adsorbed beads (belonging to the adsorbed chain) in yellow. The black arrow indicates the flow direction.}
	\label{snapshot}
\end{figure}

Based on these results, we propose one general picture of adsorbed chains during capillary infiltration. Initially, due to the relatively high velocity of the polymer melt, the adsorbed chains are rapidly stretched. As a result, more repeat units in the adsorbed chains are adsorbed on the surface, and the chains become highly oriented in the flow direction. As the velocity of polymer melt decreases, the adsorbed chains gradually relax towards their equilibrium states. Consequently, loop structures increases and adsorbed chains become less oriented, resembling the relaxation stage of chains adsorbed onto a planar surface. 
A snapshot of a system with chain length $N=100$ and capillary radius $R=10.85\,\sigma$ is shown in Figure \ref{snapshot}. 

\subsection{After Capillary Filling}
\begin{figure}[ht]
	\centering
	\includegraphics[scale=0.5]{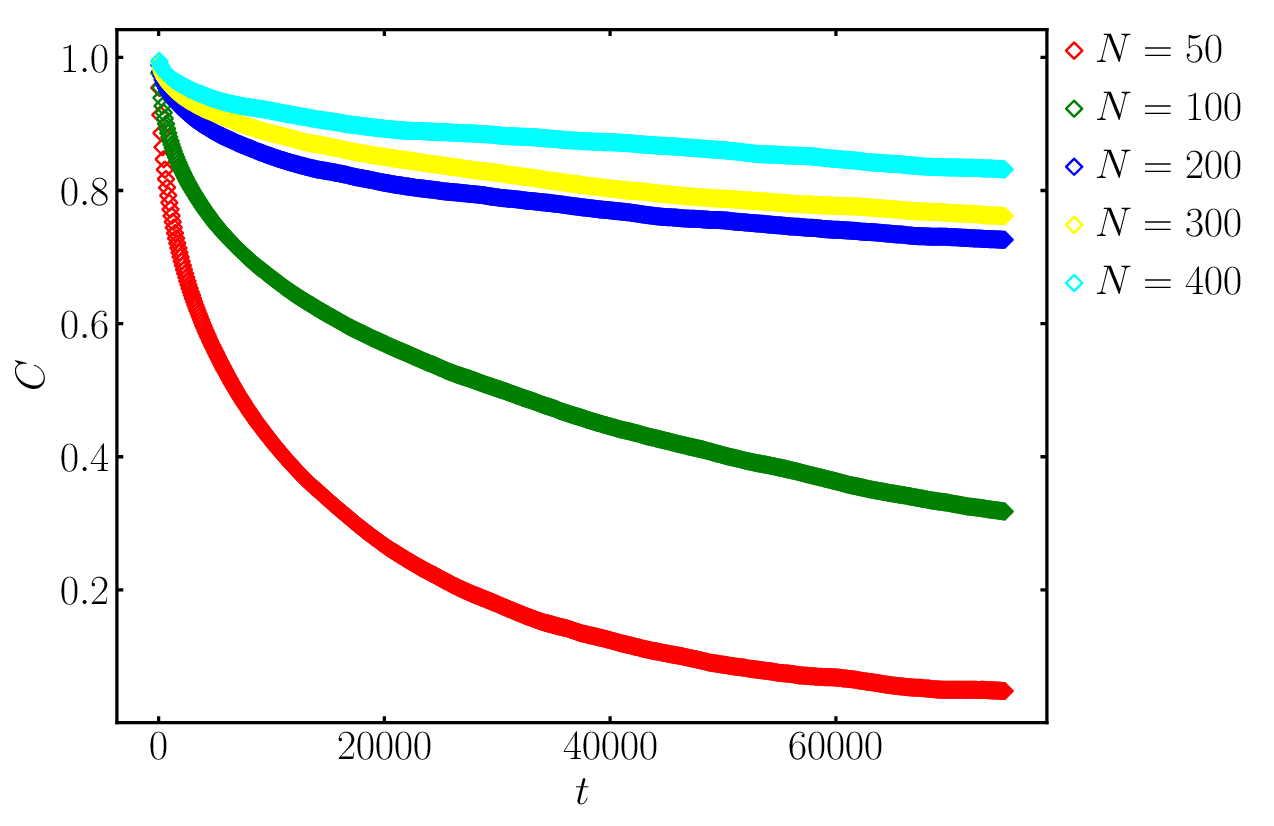}
	\caption{The autocorrelation functions of end-to-end vector in the system with tube radius $R=10.85\,\sigma$.}
	\label{RtR0_R10}
\end{figure}

After full infiltration, we explore the dynamic properties and adsorption kinetics of chains under confinement.
Firstly, we calculate the autocorrelation function of the end-to-end vector.
The relaxation dynamics of chains inside the tube can be represented by autocorrelation functions, defined as follows:
\begin{equation}
	C(t) = \frac{\langle R_tR_0\rangle}{\langle |R_t||R_0|\rangle}
\end{equation} 
where $R_t$ and $R_0$ are end-to-end vectors at a time interval $t$ and $\langle \cdot \rangle$ represents the statistical average. 
Figure \ref{RtR0_R10} shows the autocorrelation in system with a capillary tube radius $R=10.85\,\sigma$. 
The results indicate that relaxation dynamics of chains slows down as the chain length increases. However, when the chain length is sufficiently long, the variation in dynamics becomes less pronounced, which can be attributed to the strong confinement of chains inside the tube. 
Under high confinement, chains are highly oriented along the tube axis, leading to extremely slow autocorrelation delay. 
Consequently, the differences in dynamics for longer chains become less obvious.

The adsorption kinetics of repeat units and chains are investigated using the normalized survival functions of adsorbed repeat units and chains after full infiltration. 
At $t^*$ when the capillary tube is fully filled, we label the adsorbed repeat units and the adsorbed chains. 
Their amounts are expressed as $N(0)$ and $n_c(0)$, respectively. 
After a time $t$ since full infiltration, the amounts of the repeat units and chains that remain adsorbed are denoted as $N(t)$ and $n_c(t)$, respectively.
The normalized survival functions of adsorbed repeat units $N(t)/N(0)$ and  adsorbed chains $n_c(t)/n_c(0)$ can be fitted by a single exponential function \cite{Douglas_1993}, defined as:
\begin{align}
	N(t)/N(0)=\mathrm{exp}(-t/\tau_m) \\
	n_c(t)/n_c(0)=\mathrm{exp}(-t/\tau_c)
	\label{survival_function}
\end{align}
where, $\tau_m$ and $\tau_c$ are the characteristic adsorption times of repeat units and chains, respectively. 

\begin{figure}[htp]
	\centering
	\begin{subfigure}[ht]{0.5\textwidth}
		\centering
		\includegraphics[width=\textwidth]{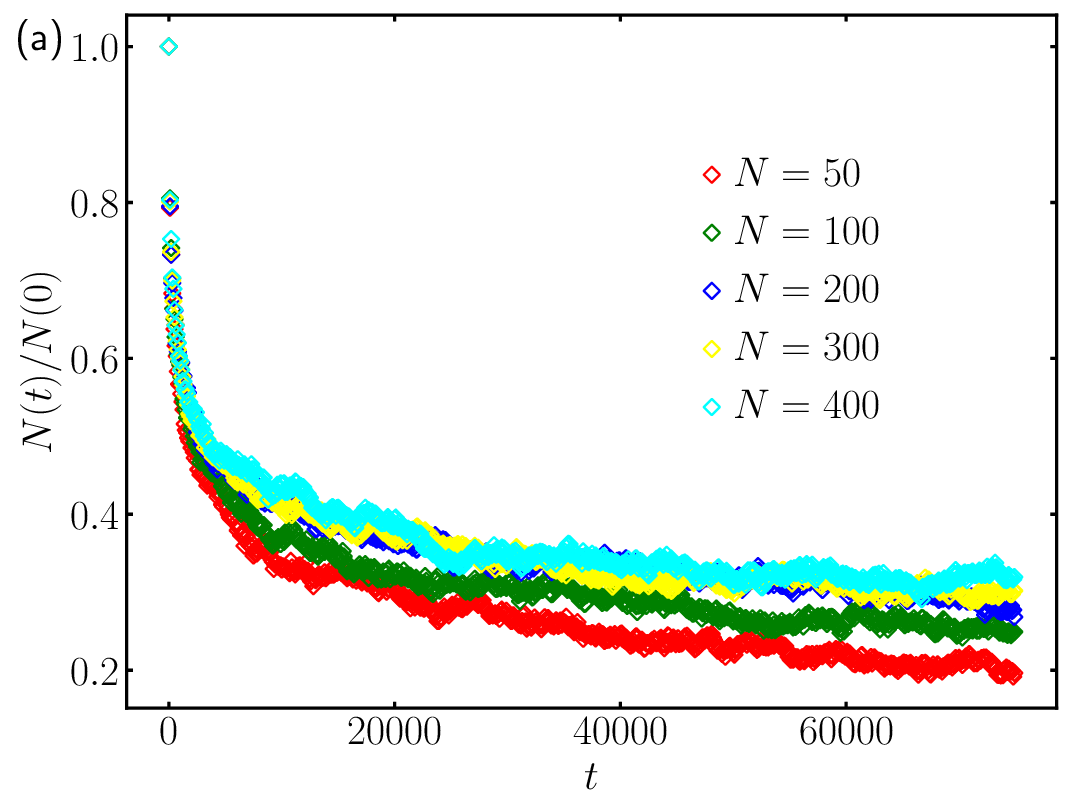}
		\label{desorption_R10_mono}
	\end{subfigure}
	\vspace{0cm}
	\begin{subfigure}[ht]{0.5\textwidth}
		\centering
		\includegraphics[width=\textwidth]{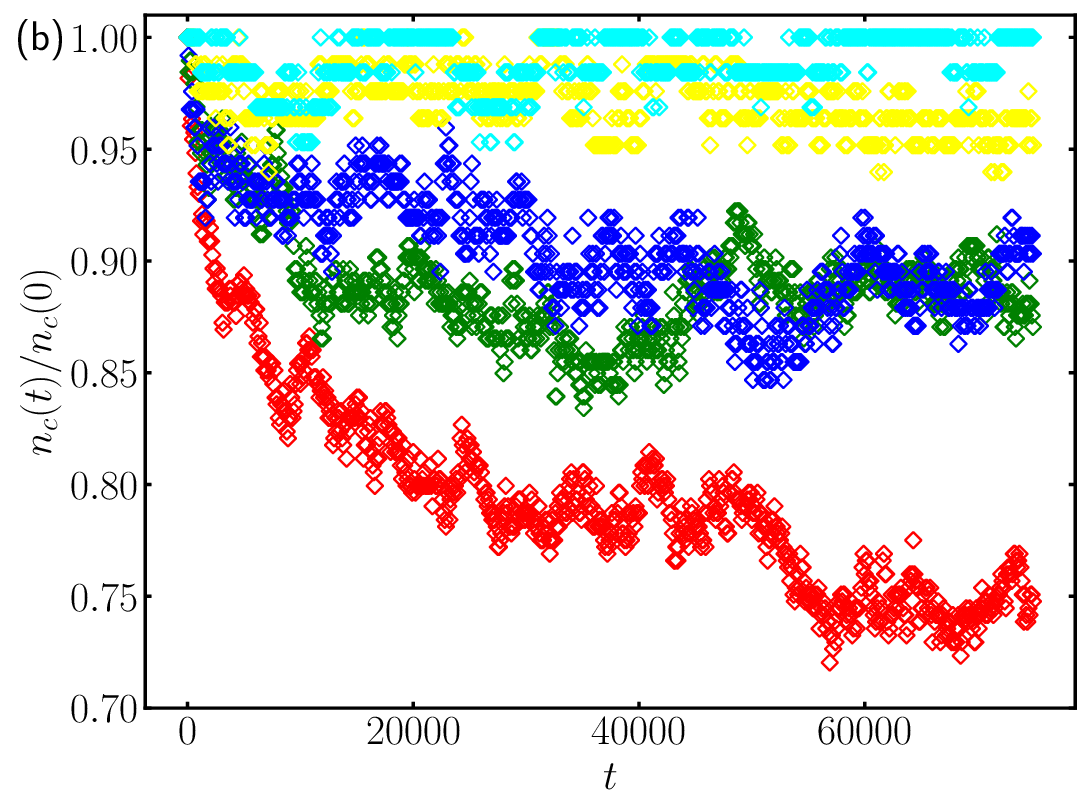}
		\label{desorption_R10_chain}
	\end{subfigure}
	\vspace{0cm}
	\caption{Normalized adsorption survival functions of adsorbed repeat units (a) and adsorbed chains (b) after fully infiltration in system with tube radius $R=10.85\,\sigma$.}
	\label{desorption_R10}
\end{figure}

We analyze the chain length dependence of adsorption kinetic using systems with a constant tube radius and different chain lengths. Figure \ref{desorption_R10} presents the results for a system with tube radius $R=10.85\,\sigma$. The survival function of repeat units indicates that repeat unit adsorption is a fast process and weakly depends on chain length, whereas chain adsorption is a slower process and strongly depends on chain length.  

\begin{figure}[htp]
	\centering
	\begin{subfigure}[ht]{0.5\textwidth}
		\centering
		\includegraphics[width=\textwidth]{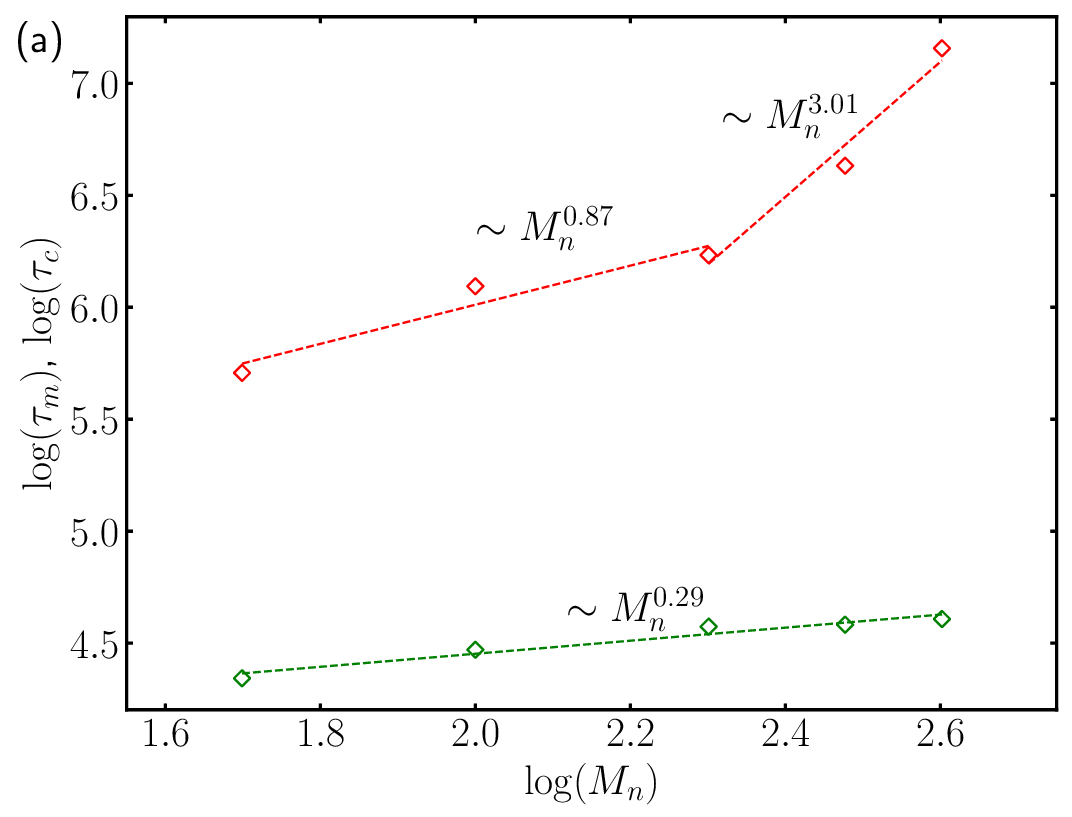}
	\end{subfigure}
	\vspace{0cm}
	\begin{subfigure}[ht]{0.5\textwidth}
		\centering
		\includegraphics[width=\textwidth]{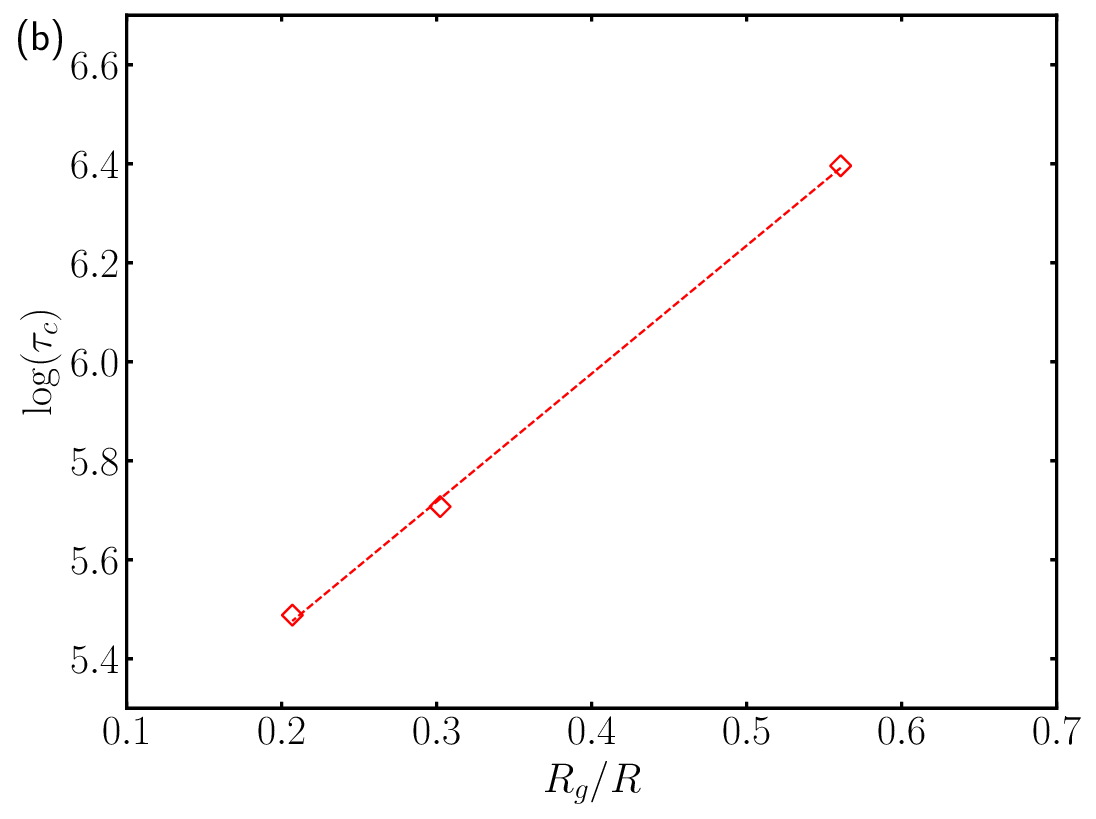}
	\end{subfigure}
	\vspace{0cm}
	\caption{(a): The chain length dependence of the desorption characteristic time for repeat units $\tau_m$ (green diamonds) and chains $\tau_c$ (red diamonds) after full infiltration. The dashed lines indicate linear fits. (b): The characteristic adsorption time of chains as a function of the degree of confinement, $R_g/R$.}
	\label{characteristic}
\end{figure}

Figure \ref{characteristic}(a) shows the result of adsorption characteristic times. For the kinetic of exchange between adsorbed and non-adsorbed chains, the scaling theory developed by de Gennes predicts the longest characteristic time scales as $\tau_c \sim \tau_mN^3$ when considering entanglement, where $\tau_m$ is the desorption time of a repeat unit. In our study, since the chain length ranges from below to above the entanglement threshold ($N_e=80$) \cite{Everaers_20, Sukumaran_2005}, we observe a cross over from $\tau_c \sim \tau_mN^{0.87}$ to $\tau_c \sim \tau_mN^{3.01}$, which is consistent with scaling theory.

To study the influence of confinement on adsorption kinetic, we study systems with a constant chain length and different tube radii, as shown in Figure \ref{characteristic}(b). The results show that the characteristic adsorption time is a function of the degree of confinement $R_g/R$, where $R_g$ is the chain's gyration radius in bulk. Thus, the characteristic adsorption time increases as the degree of confinement increases, which is consistent with the experimental results of Tu \textit{et al.} \cite{Tu_2021}.

\section{Summary}
In the first part, we explore the configurations of adsorbed chains during capillary in systems with different capillary radii and chain lengths. Firstly, we calculate the time-dependent amount of mobile repeat units inside the capillary tube during infiltration. In the system with a capillary tube radius $R=5.85\,\sigma$, the increase in mobile repeat units slows down and the fraction of mobile repeat units among all repeat units inside tube decreases during infiltration. These results are consistent with the experimental findings of dielectric strength in AAO arrays with a diameter $R=20\,\text{nm}$. In systems with capillary tube radii $R=10.85,15.85\,\sigma$, this slowing-down tendency is not observed, which aligns with the experimental results in AAO arrays with a diameter $R=50\,\text{nm}$. Thus our simulation model successfully reproduces the experimental findings. Secondly, we calculate the fractions of three structural components  (train, loop, tail) in adsorbed chains. We find that $f_{\mathrm{loop}}$ increases while $f_{\mathrm{train}}$ decreases during capillary infiltration. Additionally, the end-to-end distance of adsorbed chains in the flow direction initially increases and then decreases at later times. These results suggest that increase in $f_{\mathrm{loop}}$ is a consequence of a stretched-relaxed transition during capillary filling.

In the second part, we analyze the dynamic properties and adsorption kinetic after full infiltration. The results of autocorrelation function show that relaxation dynamics of chains inside the capillary tube decrease as the chain length increases. When chain length is sufficiently long, relaxation dynamics become extremely slow due to strong confinement inside capillary tube. The adsorption kinetic of repeat units and chains are also investigated. The characteristic desorption time of repeat units exhibits weak chain length dependence, while the characteristic desorption time of chains shows strong chain length dependence. The scaling relationship between characteristic adsorption time and chain length is consistent with the scaling theory developed by de Gennes \cite{de_Gennes_1987}. The result in a system with a constant chain length and different tube radii shows that the characteristic adsorption time increases as the degree of confinement increases. 

One possible extension of our work is to understand the influence of the adsorbed chains' conformations on capillary filling of polymers of different architectures. 
With systems of statistical copolymers \cite{Heo_2024}, non-monotonic speed of imbibition was obeserved when the fraction of one type of monomer changes.
Diblock copolymer presented another interesting subject.
A more comprehensive understanding of adsorbed chains, including their interactions with free chains and the influence of chain orientation \cite{Jeong_2017}, could significantly enhance our understanding of the dynamics underlying capillary imbibition.

\section{Supplementary Material}
See the supplementary material for detailed simulation model, supplementary result and discussion.

\begin{acknowledgments}
This research was supported the National Key R\&D Program of China (Grant No. 2022YFE0103800), the National Natural Science Foundation of China (Grant No. 22373036), R\&D Program of Guangzhou (Grant No. 2024D03J0007), and KingFa Science and Technology Company. 
The computation of this work was made possible by the facilities of Information and Network Engineering and Research Center of SCUT.
\end{acknowledgments}

\bibliography{capillary_cite}

\clearpage
\setcounter{section}{0}


\title{Chain Conformations in Adsorbed Layer during Polymer Capillary Imbibition}

\author{Tao Liang}
\affiliation{South China Advanced Institute for Soft Matter Science and Technology, School of Emergent Soft Matter, South China University of Technology, Guangzhou 510640, China}
\affiliation{Guangdong Provincial Key Laboratory of Functional and Intelligent Hybrid Materials and Devices, South China University of Technology, Guangzhou 510640, China}

\author{Jiajia Zhou}
\affiliation{South China Advanced Institute for Soft Matter Science and Technology, School of Emergent Soft Matter, South China University of Technology, Guangzhou 510640, China}
\affiliation{Guangdong Provincial Key Laboratory of Functional and Intelligent Hybrid Materials and Devices, South China University of Technology, Guangzhou 510640, China}



\renewcommand{\thefigure}{S\arabic{figure}}
\setcounter{figure}{0}
\begin{center}
	{\large \textbf{Supplementary Information}}
\end{center}
\begin{center}
	{\large \textbf{Chain Conformations in Adsorbed Layer during Polymer Capillary Imbibition}}
\end{center}



\section{Simulation Model}
\begin{figure}[ht]
	\centering
	\begin{subfigure}[ht]{0.4\textwidth}
		\centering
		\includegraphics[width=\textwidth]{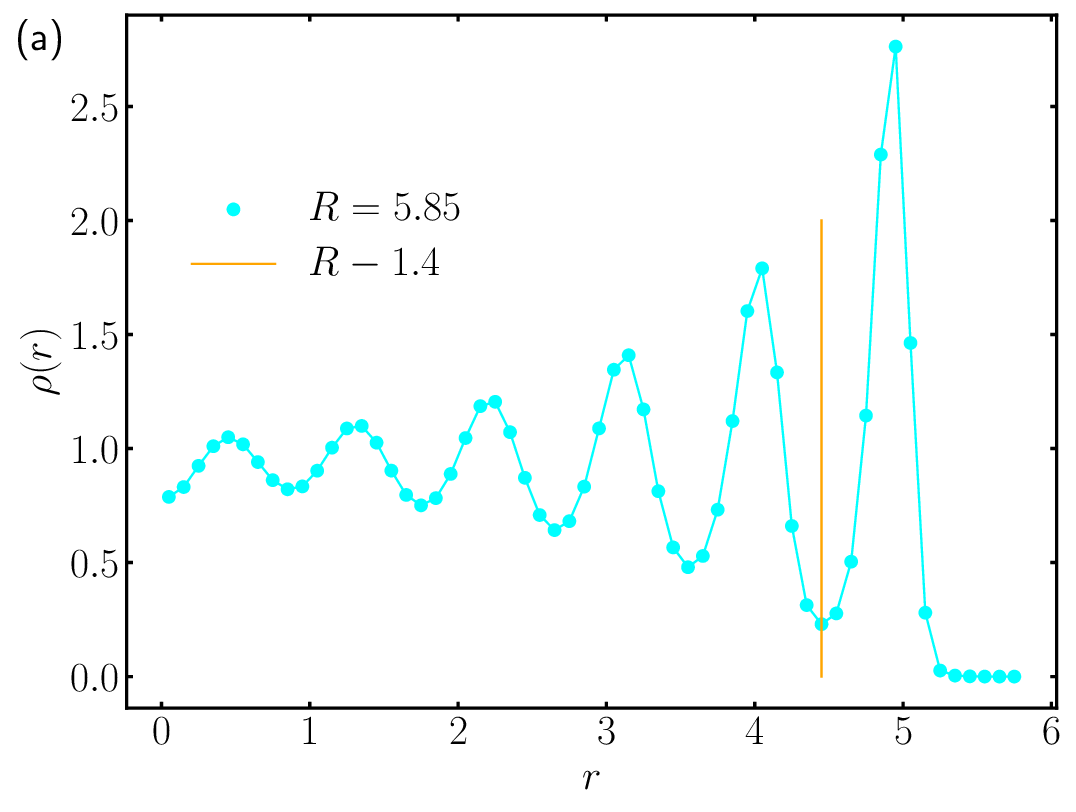}
		\label{density_a}
	\end{subfigure}
	\vspace{0cm}
	\begin{subfigure}[ht]{0.4\textwidth}
		\centering
		\includegraphics[width=\textwidth]{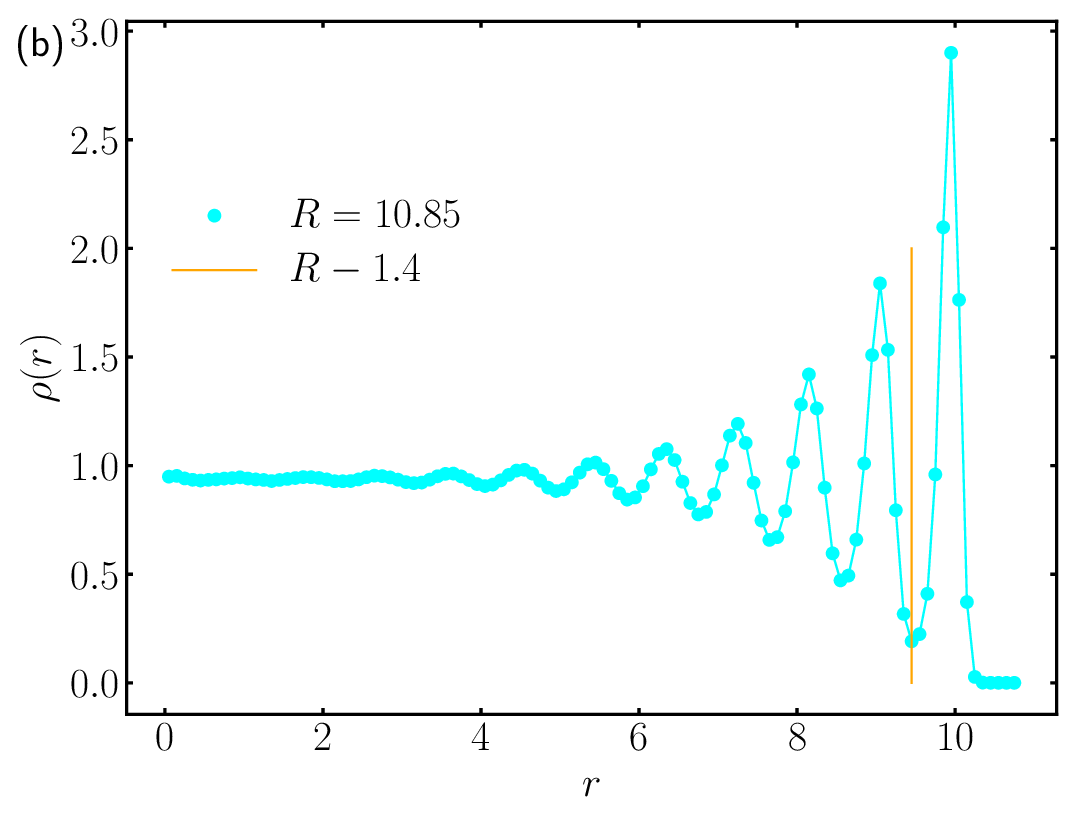}
		\label{density_b}
	\end{subfigure}
	\vspace{0cm}
	\begin{subfigure}[ht]{0.4\textwidth}
		\centering
		\includegraphics[width=\textwidth]{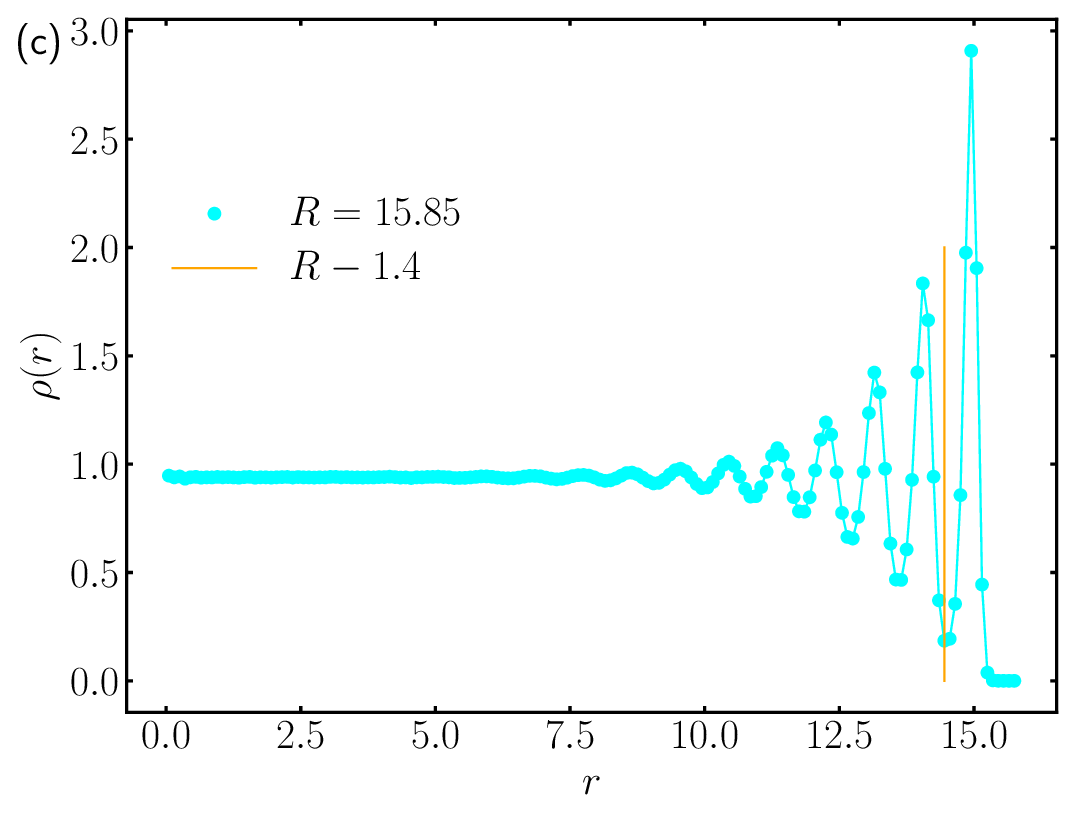}
		\label{density_c}
	\end{subfigure}
	\vspace{0cm}
	\caption{The density profile of monomers as a function of distance $r$ from the tube axis with radius $R=5.85\,\sigma$ (a), $R=10.85\,\sigma$ (b), $R=15.85\,\sigma$ (c). The orange lines indicate the valleys between first two layers from surface.}
	\label{density}
\end{figure}

The density profiles of monomers as a function of distance $r$ from the capillary tube axis with three radii $R=5.85, 10.85, 15.85\,\sigma$ and chain length $N=50$. The density profiles are drawn after full infiltration. As shown in Figure \ref{density}, monomers near inner surfaces exhibit a layered distribution. The valleys between the first two layers in three systems are all located in $R-1.4\,\sigma$. 

\begin{figure}[H]
	\centering
	\includegraphics[scale=0.5]{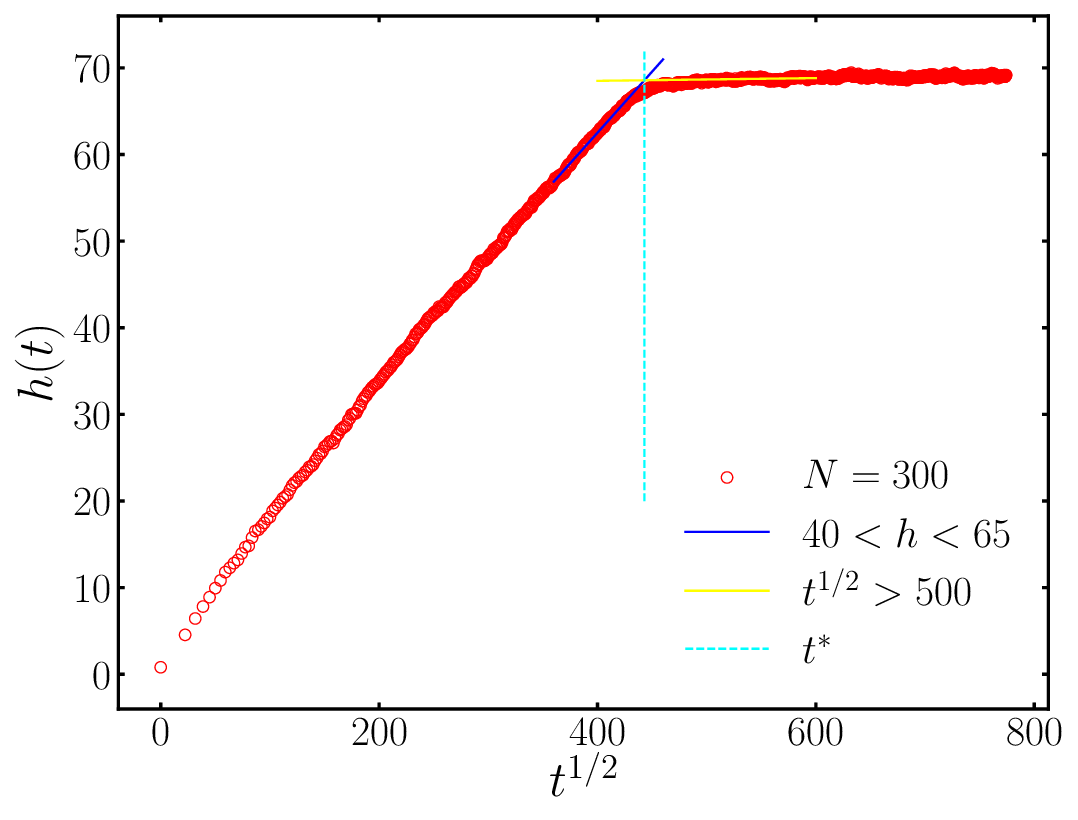}
	\caption{The capillary infiltration height $h(t)$ as a function of $t^{1/2}$. Blue solid line is fitted based on linear function from $h\in(40, 65)$, yellow line is fitted from $t^{1/2}>500$. Vertical cyan dash line locates at the intersection point of two solid lines.}
	\label{R10N300h_t}
\end{figure}

Figure \ref{R10N300h_t} shows the time-dependence of infiltration height $h(t)$ as a function of square root time. Two lines are fitted based on linear function from $h\in(40, 65)$ and $t^{0.5}>500$, respectively. The intersection point of two lines indicates the time $t^*$ when the capillary tube is fully filled.
\newpage
\section{Result and Discussion}
\begin{figure}[htp]
	\centering
	\includegraphics[scale=0.5]{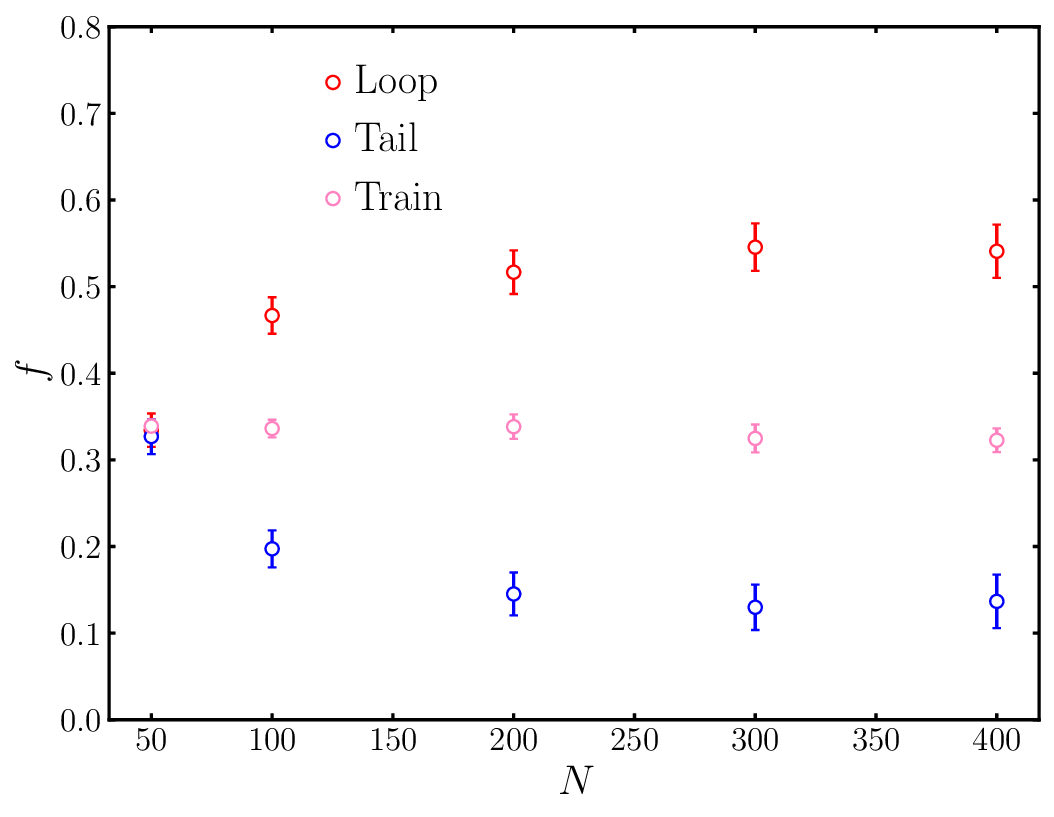}
	\caption{The ratio of three kinds of repeat units in equilibrium state in system with tube radius $R=5.85\,\sigma$ as chain lengths increase.}
	\label{ratio_loop_N}
\end{figure}
\begin{figure}[H]
	\centering
	\includegraphics[scale=0.4]{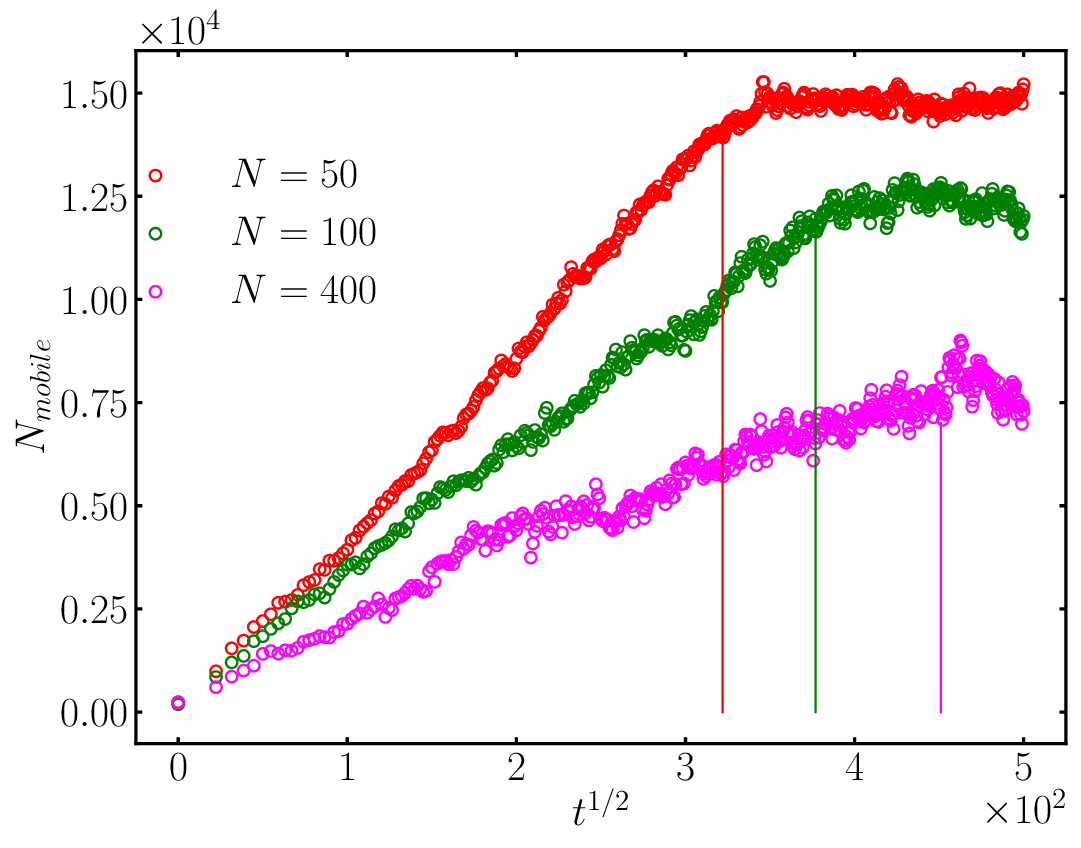}
	\caption{The amount of mobile monomers versus $t^{1/2}$ with capillary radii $R=10.85\sigma$, polymer melt of length $N=50$ is shown in red, $N=100$ is shown in green and $N=400$ is shown in purple. The vertical solid  lines indicate the time $t^*$ when capillary is full filled.}
	\label{mobile_10}
\end{figure}
\begin{figure}[H]
	\centering
	\includegraphics[scale=0.4]{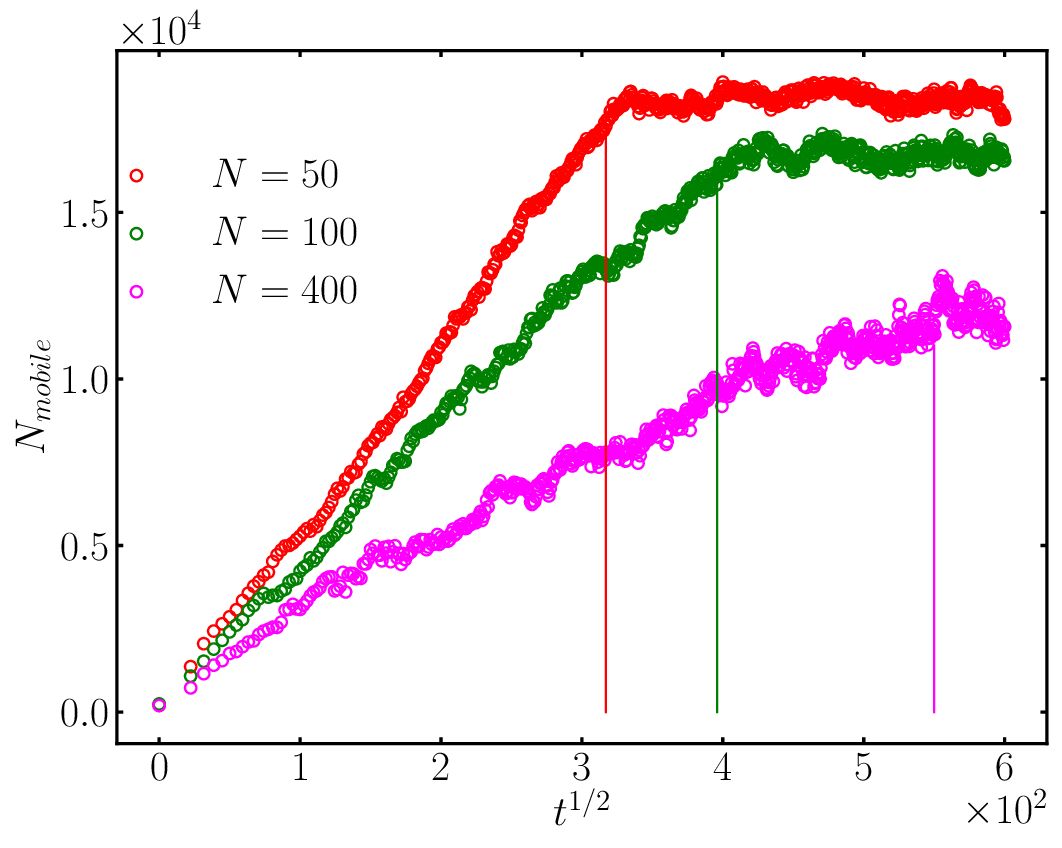}
	\caption{The amount of mobile monomers versus $t^{1/2}$ with capillary radii $R=15.85\sigma$, polymer melt of length $N=50$ is shown in red, $N=100$ is shown in green and $N=400$ is shown in purple. The vertical solid  lines indicate the time $t^*$ when capillary is full filled.}
	\label{mobile_15}
\end{figure}
\begin{figure}[H]
	\centering
	\includegraphics[scale=0.5]{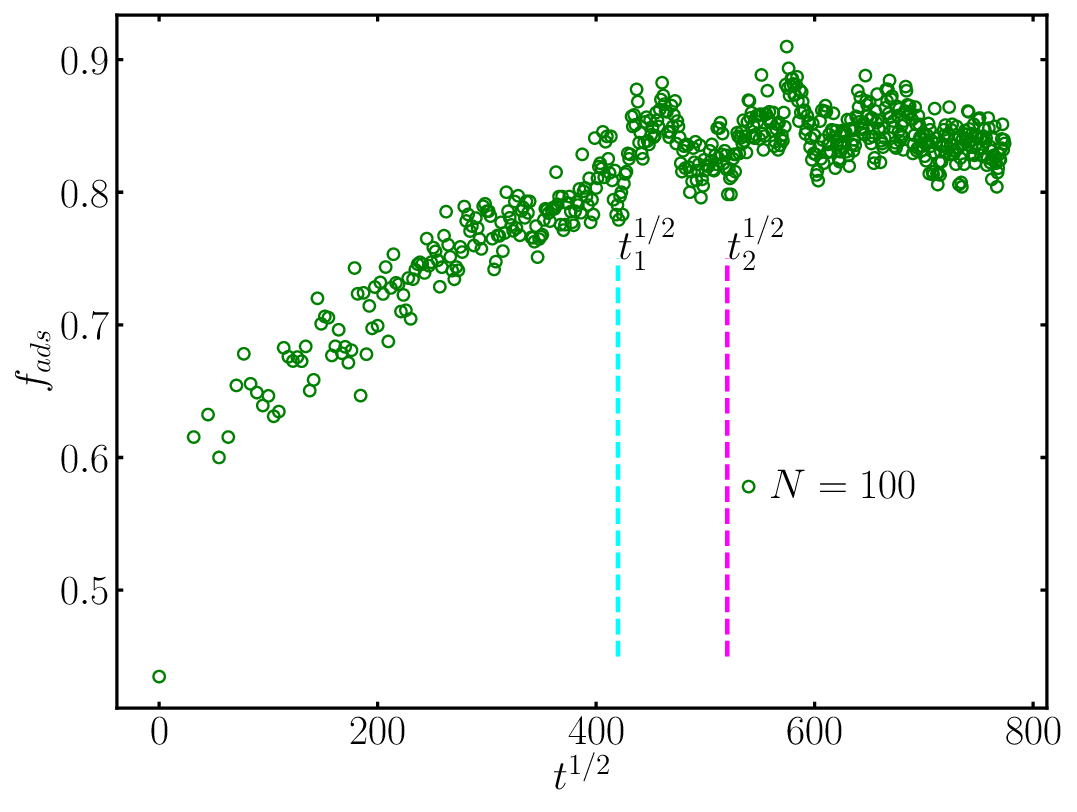}
	\caption{The ratio between the amount of adsorbed chains and the amount of chains that are inside the tube in system with tube radius $R=10.85\,\sigma$ and chain length $N=100$. Two dashed lines indicate $t_1^{1/2}=420$, $t_2^{1/2}=520$, respectively.}
	\label{r_ads_R10}
\end{figure}
\begin{figure}[htp]
	\centering
	\begin{subfigure}[ht]{1.0\textwidth}
		\centering
		\includegraphics[width=\textwidth]{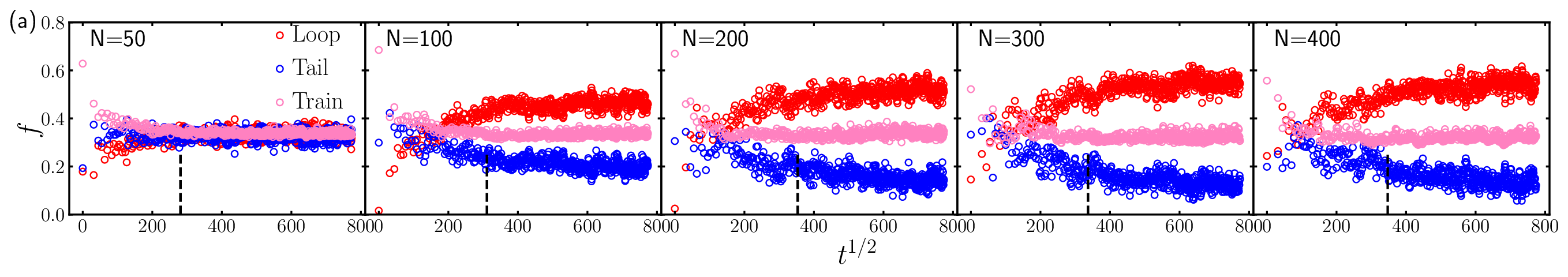}
	\end{subfigure}
	\vspace{0cm}
	\begin{subfigure}[ht]{1.0\textwidth}
		\centering
		\includegraphics[width=\textwidth]{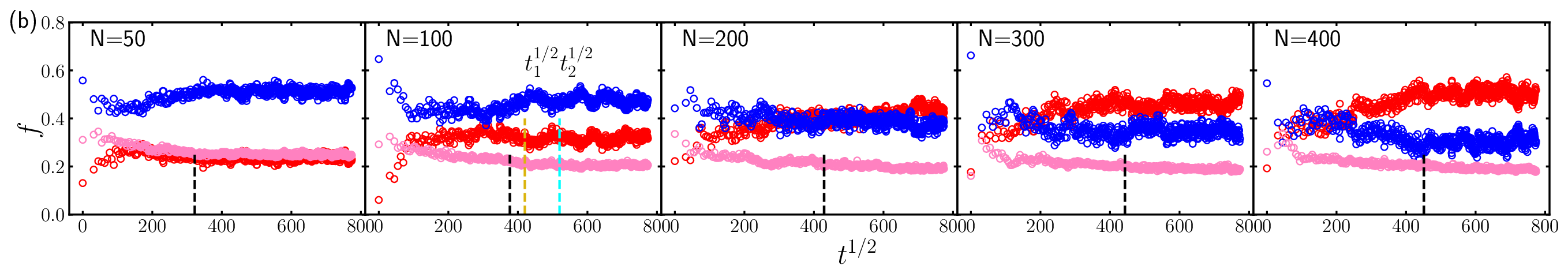}
	\end{subfigure}
	\vspace{0cm}
	\begin{subfigure}[ht]{1.0\textwidth}
		\centering
		\includegraphics[width=\textwidth]{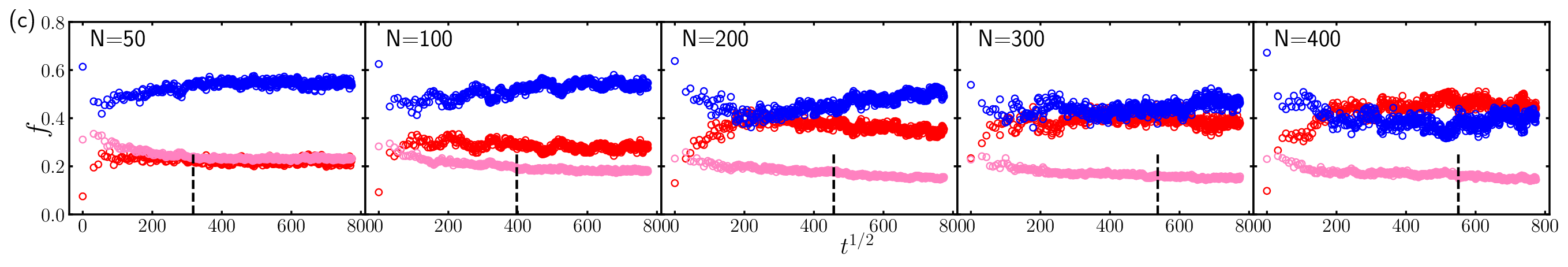}
	\end{subfigure}
	\vspace{0cm}
	\caption{The ratios of three types repeat units (loop shown in red, train shown in pink and tail shown in blue) in adsorbed layer with capillary tube radii $R=5.85\sigma$ (a), $R=10.85\sigma$ (b), $R=15.85\sigma$ (c). The vertical black dashed lines indicate the time when capillary tube is fully filled. In system with $R=10.85\,\sigma$ and $N=100$, two colored dashed lines indicate $t_1^{1/2}=420$, $t_2^{1/2}=520$, respectively.}
	\label{ratio}
\end{figure}

The fraction of mobile monomers among all monomers inside the capillary tube during infiltration process with capillary radius $R=10.85\,\sigma$ is shown in Figure \ref{mobile_10}. The fraction of mobile monomers among all monomers inside the capillary tube with capillary radius $R=15.85\,\sigma$ is shown in Figure \ref{mobile_15}.
We calculate the ratio $f_{\mathrm{ads}}$ between the amount of adsorbed chains and the amount of all chains inside the tube in system with tube radius $R=10.85\,\sigma$ and chain lengths $N=100$. The result in Figure \ref{r_ads_R10} shows that $f_{\mathrm{ads}}$ increases with time. We label two time points which indicate two times when $f_{\mathrm{ads}}$ increase rapidly. Surprisingly, we can observe the decrease of $f_{\mathrm{loop}}$ in the same times in Figure \ref{ratio}. We could imagine that one free chain would contribute a large tail ratio and a small loop ratio if just one end gets adsorbed on the tube surface. That is why at the two labeled times, average loop ratio decrease and tail ratio increase.
The ratios of three types monomers of adsorbed chains in systems with all capillary radii $R=5.85,10.85,15.85\,\sigma$ and all chain lengths $N=50,100,200,300,400$ are shown in Figure \ref{ratio}. Although there are some differences, an increase in the ratio of loop monomers $f_{\mathrm{loop}}$ and a decrease in the ratio of train monomers $f_{\mathrm{train}}$ can be observed in all systems.  

\begin{figure}[H]
	\centering
	\begin{subfigure}[H]{0.5\textwidth}
		\centering
		\includegraphics[width=\textwidth]{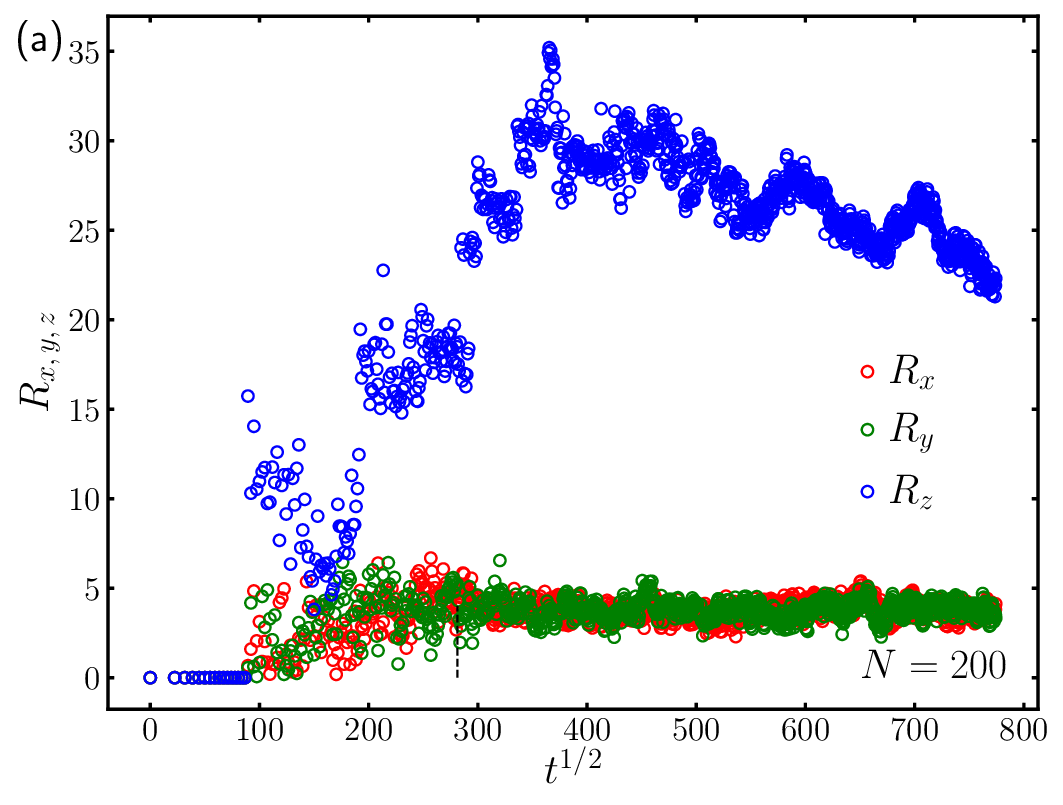}
	\end{subfigure}
	\vspace{0cm}
	\begin{subfigure}[H]{0.5\textwidth}
		\centering
		\includegraphics[width=\textwidth]{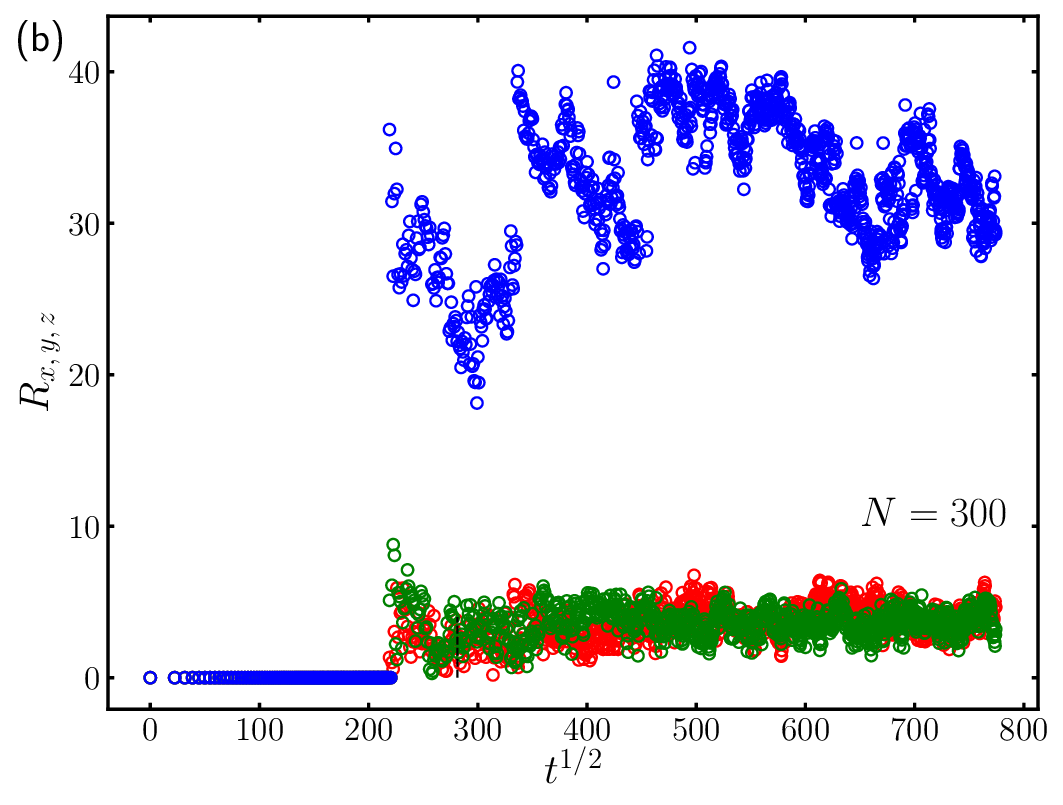}
	\end{subfigure}
	\vspace{0cm}
	\begin{subfigure}[H]{0.5\textwidth}
		\centering
		\includegraphics[width=\textwidth]{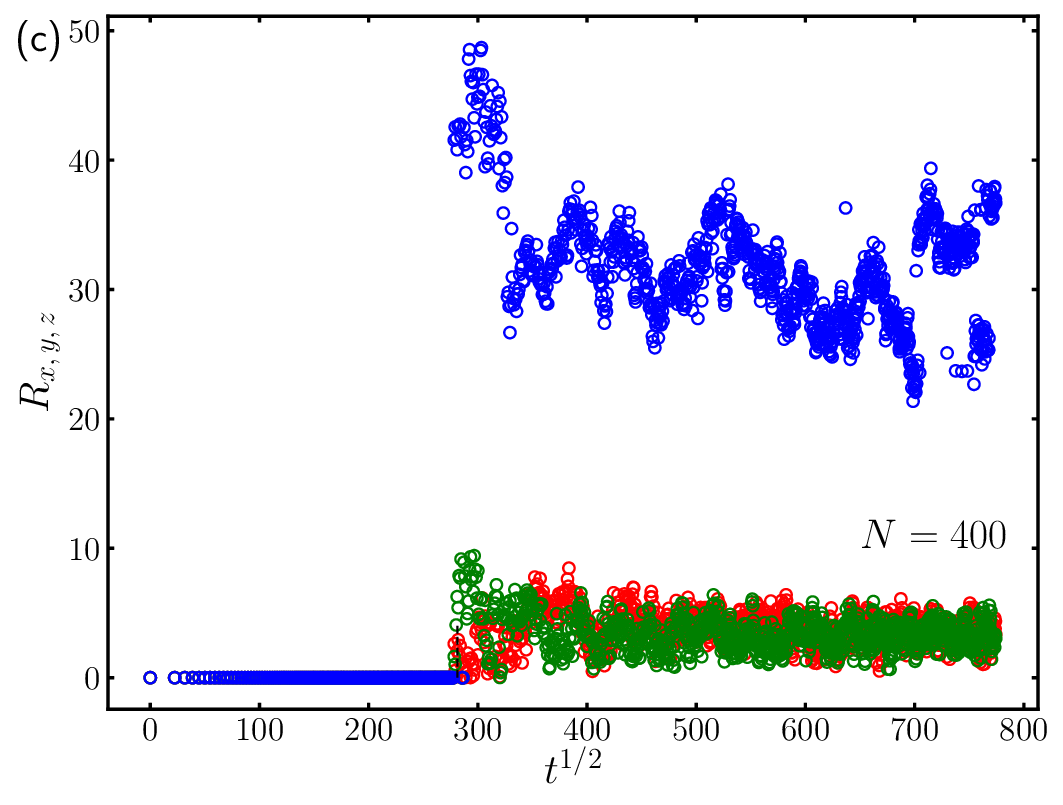}
	\end{subfigure}
	\caption{The end-to-end distances in \textit{x, y, z} directions $R_{x,y,z}$ during capillary infiltration with tube radius $R=5.85\,\sigma$. Chain length $N=200$ (a), $N=300$ (b), $N=400$ (c). The vertical dash lines indicate the time when capillary tube is fully filled.}
	\label{R5xyz_200300400}
\end{figure}

\clearpage
Figure \ref{R5xyz_200300400} shows the results of end-to-end distances in (\textit{x, y, z}) directions $R_{x,y,z}$ during capillary infiltration in systems with tube radius $R=5.85\,\sigma$ and chain lengths $N=200, 300, 400$.

\end{document}